\documentclass[%
 reprint, amsmath, amssymb, aps,]{revtex4-1}
\usepackage{booktabs}
\usepackage{graphicx}
\usepackage{dcolumn}
\usepackage{bm}
\usepackage{color}
\usepackage{url}
\usepackage[normalem]{ulem}
\usepackage{mathtools}
\usepackage{array}
\newcolumntype{P}[1]{>{\centering\arraybackslash}p{#1}}
\newcolumntype{M}[1]{>{\centering\arraybackslash}m{#1}}
\usepackage[caption=false]{subfig}
\usepackage{dcolumn}
\usepackage{graphicx, epsfig}
\usepackage[dvipsnames]{xcolor}
\usepackage{mathrsfs}
\usepackage{yhmath}
\usepackage[caption=false]{subfig}
\usepackage[normalem]{ulem}
\usepackage{mathtools}
\usepackage{bigints}
\usepackage{float}
\usepackage[colorlinks = true, linkcolor = purple, urlcolor  = blue, citecolor = blue, anchorcolor = blue]{hyperref}
\usepackage{float}
\usepackage{multirow}

\newcommand{\RN}[1]{%
  \textup{\uppercase\expandafter{\romannumeral#1}}%
}

\def\lsim{\;\raise0.3ex\hbox{$<$\kern-0.75em\raise-1.1ex\hbox{$\sim$}}\;}
\def\gsim{\;\raise0.3ex\hbox{$>$\kern-0.75em\raise-1.1ex\hbox{$\sim$}}\;}

\def\cmc{\rm ~cm^{-3}}
\def\kms{\rm ~km~s^{-1}}

\def\cmc{\rm ~cm^{-3}}
\def\diff{\rm ~cm^2~s^{-1}}
\def \kms {\rm ~km~s^{-1}}

\def\ergs{\rm ~erg~s^{-1}}

\def\enf{\rm ~erg~cm^{-2}~s^{-1}}

\def\ecsb{erg cm$^{-2}$ s$^{-1}$ arcsec$^{-2}$ }
\def\ecsb2{erg cm$^{-2}$ s$^{-1}$ arcsec$^{-2}$}

\def\V4641~Sgr {V4641~Sgr}

\def\arcmin{\hbox{$^\prime$}}
\def\arcsec{\hbox{$^{\prime\prime}$}}
\def\apj{ApJ}
\def\mnras{MNRAS}
\def\nat{Nat}

\def\araa{ARA\&A}                
\def\aap{A\&A}                   
\def\apjs{ApJS}                  
\def\apjl{ApJ}                   

\def\ssr{Space Sci. Rev.}

\def\apj{ApJ}
\def\apjl{ApJ Lett.}
\def\mnras{MNRAS}
\def\nat{Nature}

\def\prd{Phys. Rev. D}

\def\araa{Ann. Rev. Astron. Astrophys.}                
 \def\aap{Astron. Astrophys.}                  
\def\apjs{Astrophys. J. Suppl. Ser.}                  
\def\apjl{ApJ Lett.}                   

\def\ssr{Space Sci. Rev.}

\usepackage{url}
\usepackage{xcolor}
\definecolor{newcolor}{rgb}{.8,.349,.1}

\begin{document}

\preprint{APS/123-QED}




\title{Particle acceleration and non-thermal broadband spectrum of \\
extended shocked outflows in black hole binary V4641~Sgr - a galactic pevatron}%

\author{A. M. Bykov$^1$} 
\email{byk@astro.ioffe.ru}
\author{S. M. Osipov$^1$} 
\email{osm.astro@mail.ioffe.ru}
\author{V. I. Romansky$^1$}
\email{romanskyvadim.astro@mail.ioffe.ru}
\author{E. M. Churazov$^{2,3}$} 
\email{churazov@mpa-garching.mpg.de}
\author{I. I. Khabibullin$^{4,2,3}$} 
\email{ildar@mpa-garching.mpg.de}


\address{$^1$Ioffe Institute, Saint-Petersburg, Polytechnicheskaya str., 26, 194021, Russia}
\affiliation{$^2$Max Planck Institute for Astrophysics, Karl-Schwarzschild-Str. 1, D-85741 Garching, Germany}
\affiliation{$^3$Space Research Institute (IKI), Profsoyuznaya 84/32, Moscow 117997, Russia}
\affiliation{$^4$Rudolf Peierls Centre for Theoretical Physics, Department of Physics, University of Oxford, Clarendon Laboratory, Parks Rd, Oxford, OX1 3PU, United Kingdom}


\begin{abstract} 
 Stellar mass black holes accreting at a super-Eddington rate produce strong outflows and non-thermal radiation up to PeV energy. Recently,  very high energy (VHE) emission was discovered from the environment of the galactic microquasar  V4641 Sgr, which is known to experience super-Eddington outbursts. We present MHD simulations supplemented with a nonlinear kinetic Monte Carlo model that reproduces the observed radio, X-ray, and gamma-ray emission from the region over tens of parsecs surrounding V4641 Sgr. The reported multi-wavelength non-thermal emission can be explained as produced by PeV electrons accelerated by a system of strong shock waves created by the interaction between a narrow trans-relativistic polar jet and a quasi-spherical accretion-disk wind. The shock positions are associated with bright spots of the gamma-ray and  X-ray radiation, whose spatial profiles can be tested with dedicated observations. The model predicts that a sizable fraction of the jet's kinetic energy is channeled into TeV-PeV protons, 
suggesting
that such systems might be powerful sources of PeV-range cosmic rays during the accelerator activity cycle. In this model, the bow-tie shape structure discovered by MeerKAT in V4641 Sgr is associated 
with the radio emission from the termination surface of the quasi-spherical accretion-disk wind.  The newly presented (from the SRG/eROSITA all-sky survey) and archival (INTEGRAL and Swift BAT) X-ray data are quantitatively consistent with the model, and simultaneously reproduce VHE measurements with H.E.S.S., HAWC, and LHAASO. Deeper X-ray observations will provide a test of the intermittent operation of the accelerator, in contrast with more persistent sources such as SS433/W50.
\end{abstract}

\pacs{Valid PACS appear here}
\maketitle

\section{Introduction}

Recent sensitive observations of very high energy photon sources revealed that galactic gamma-ray binaries can accelerate particles above PeV energies. This may help to resolve the long-standing problems of the origin of galactic cosmic rays (CRs) in the energy range above the PeV spectral knee. Ultra high energy photons were detected from the known galactic stellar mass black holes accreting at about the critical rate, which are the microquasars SS~433, V4641~Sgr, GRS 1915+105, MAXI J1820+070  \citep{V4641_HAWC_2024Natur.634..557A,LHAASO_MQs_2025NSRev..12af496L, V4641_HESS_2026A&A...706A...8A}, Cyg X-3 \citep{CygX3_LHAASO25}.      
These gamma-ray observations made the microquasars promising candidates for galactic ultra high energy particle accelerators.

The detected very high energy (VHE) photons can be produced by both leptonic and hadronic radiation mechanisms \citep[see e.g.][]{2004vhec.book.....A,2005A&A...432..609B, 2010MNRAS.404L..55D,2014SSRv..183..371P, 2017SSRv..207....5R,2026JHEAp..5100538C}  widely discussed in the recent interpretation of microquasar PeV regime gamma-ray fluxes \citep[see e.g.][]{2024arXiv241108762P,2024JHEAp..43...93K,2025PhRvD.111j3025N,V4641_lepton_2025arXiv251213578K,shear2026ApJ...997..128W,V4641_bin_accr26}. The VHE observations unequivocally proved the presence of PeV energy particles, but the question of particle acceleration mechanism remains open. 

Different mechanisms of high energy particle acceleration by turbulence inside hot accretion flows in the close vicinity of the accreting object \citep[see e.g.][]{2019MNRAS.485..163K,2025A&A...697A.124L} and within the parsec-scale extended nebula \citep{2025arXiv251202574D,2026arXiv260309394D} were discussed. The magnetic reconnection processes above the super-critically accreting black hole are considered in \citet[][]{2026APh...17703214P}.  \citet{V4641_lepton_2025arXiv251213578K} suggested that particle acceleration occurs in vicinity of the black hole.
In this model, the spectra and morphology of the X-ray and VHE radiation around the microquasar are determined by the propagation of injected relativistic particles (protons and electrons) along the galactic magnetic field in the interstellar medium.    

 Another class of models suggested the presence of extended outflows of the superluminous microquasars well away from the accreting black hole. The acceleration of the VHE particles in the jets of microquasars by internal shocks was discussed in \citet[][]{2000A&A...356..975K}, while the jet recollimation shocks were studied as VHE particle acceleration sites in \citet[][]{Churazov,HESS2024SS433,SS433_min_model_2025PhRvD.112f3017B,2024arXiv241108762P,2026arXiv260316647P}. Also, relativistic particle re-acceleration in the jet-cocoon shear flows \citep[][]{2025PhRvD.112l3015Z,shear2026ApJ...997..128W,V4641_bin_accr26} may operate.

 The extended nature of the observed very high energy emission from SS~433 and V4641~Sgr raises the question of whether the particle acceleration site is in close vicinity of the accreting object or is located far away from the central source and is associated with its powerful parsec-scale outflows.

 The morphology of the observed extended X-ray jets as well as all the spatial and spectral characteristics of the bright knots in W50/SS~433 can be understood in the "minimalist" model of the accreting black hole outflow, which consists of a fast, extended narrow jet of velocity above 0.1 c and a broad wind of a slower speed of a few thousand $\kms$ \citep{Churazov,SS433_min_model_2025PhRvD.112f3017B}. In this model, the positions of the bright non-thermal X-ray knots are associated with strong normal shocks in the jet formed in the region of jet collision with the broad wind termination shock. The shocks are formed due to the fast flow recollimations in the jet \citep[see e.g.][]{KomissarovFalle1997}. Indeed, the initial overpressure in the jet with respect to the ambient plasma drives a number of flow recollimation shocks, either normal or oblique \citep[see e.g.][]{recoll_1997ApJ...479..151M}. Particles can be efficiently accelerated by the fast normal shock to the TeV-PeV energy regime needed to produce synchrotron X-ray radiation.

{Indeed, in the MHD simulations that we discuss below, various types of shocks and discontinuities are seen. Given the above arguments, the most promising sites of the efficient VHE particle accelerators are the strong normal shocks, which we identify by the magnitude of the entropy jump. In such shocks, efficient particle acceleration is accompanied by strong amplification of the fluctuating magnetic field due to cosmic-ray-driven instabilities, even in a jet with relatively low magnetization.}

 Deep X-ray observations of the non-thermal knots, including the X-ray polarimetry, available for W50/SS~433  are crucial to test PeV particle acceleration models. The profiles of X-ray emission of the knots e1 and e2 in W50/SS~433 \citep{Brinkmann2007,Safi-Harb2022} and especially the position angles and degree of X-ray polarization \citep{IXPE2024SS433} were reproduced in the model of particle acceleration by PeV energy by strong shocks with magnetic field amplification, which produce the observed VHE radiation in an initially low magnetized transrelativistic jet \citep{SS433_min_model_2025PhRvD.112f3017B}. The inverse Compton radiation of the accelerated electrons can completely account for the observed gamma-ray emission in this model.   
 
 The reported fluxes and the maximal energies of the detected VHE photons from V4641~Sgr are a few times higher, and its spectral index is harder than that of SS~433. This motivates an in-depth study of radio and X-ray emission from the extended region (spanning tens of parsecs) surrounding the accreting black hole in the V4641 Sgr system, with the aim of constraining possible models.

 VLA radio observations of V4641~Sgr recently analyzed by \citet{Marti_V4641_VLA_jets26} indicated that the axis of the radio jet (of about 10$\arcsec$ scale size) is likely aligned with the axis of the extended elongated structures of VHE emission above 25 TeV (of a degree scale size) detected by HAWC \citep{V4641_HAWC_2024Natur.634..557A}, LHAASO \citep{LHAASO_MQs_2025NSRev..12af496L} and H.E.S.S. \citep{V4641_HESS_2026A&A...706A...8A}. The MeerKAT L-band radio observations revealed an extended structure, which has a bow-tie shape of scale size $\sim$ 35 pc in diffuse radio emission around V4641~Sgr  \citep{MeerKaT2026A&A...706A.283G}. The authors suggested that the observed large size bow-tie structure was produced by the outflows from V4641~Sgr. 
Both results are consistent with the assumption of the minimal model of the origin of the extended  VHE radiation, as was suggested for the W50/SS~433 nebula, where, contrary to the V4641~Sgr case, no direct observational evidence for the large scale slow wind was known. The outflow velocities in V4641~Sgr measured with optical spectroscopy of broad emission line wings by \citet{V4641wind18} were $\sim$ 3000 $\kms$, and in several of the wind detections, the authors also found an active radio jet.   

 Regular X-ray monitoring of the transient superluminous X-ray binary V4641~Sgr discovered by BeppoSAX \citep{2000A&A...357..520I} and RXTE observatories in 1999 detected multiple bright events of a few-day duration associated with the black hole activity, including the super-Eddington outburst in September 1999 \citep[see, e.g.,][]{2002A&A...391.1013R}. The X-ray line spectra detected with high-resolution Chandra transmission gratings during an outburst of  V4641~Sgr, combined with optical measurements during previous outbursts, indicated that they originated in an accretion disk wind, possibly with a quasi-spherical symmetry \citep{V4641_wind_chandra2022MNRAS.516..124S}. Extended X-ray emission around  V4641~Sgr detected in September 2024 was reported by the X-Ray Imaging and Spectroscopy Mission (XRISM) \citep{2025ApJ...978L..20S}. They estimated the 
integrated X-ray flux of $(4-6)\times 10^{-12} \enf$ (2–10 keV) emission from a region of spatial extent with a radius of $(7\pm 3)'$ (i.e., 13 $\pm$ 5 pc at a distance of 6.2 kpc) assuming a Gaussian-like radial distribution. The authors suggested that the source of VHE $\sim$ 100 TeV energy electrons producing the synchrotron X-ray radiation is in the vicinity of a radius  $\lsim$ 10 pc from the microquasar,  which is much closer than the distance to the peak of VHE gamma-ray emission that was assumed to be of hadronic origin. This implied either the diffusion coefficient of the 100 TeV electrons to be suppressed to $10^{27} \diff$ or the magnitude of the magnetic field is $\sim$ 80 $\mu$G within the studied region \citep{2025ApJ...978L..20S}. The source of accelerated particles was not specified in this study.

Since the microquasars by now are among a very few potential galactic PeVatrons, it is important to estimate the contribution of the galactic microquasars to the observed population of PeV regime cosmic rays. To address this issue, detailed modeling of the formation of VHE particle spectra to V4641~Sgr is needed to study the acceleration efficiency of CR, consistent with the observational data available for the system. While the galactic microquasar SS~433 apparently accrete persistently at a highly super-critical rate \citep[e.g.][]{2021MNRAS.506.1045M}, the accretion scenario for V4641~Sgr, which has low quiescent X-ray luminosity but shows super Eddington flares \citep{2002A&A...391.1013R,2002A&A...385..904R,V4641wind18} is different, with likely high obscuration effects  \citep[see e.g.][]{2020A&A...639A..13K}. It is still a subject of ongoing studies \citep[see e.g.][]{V4641_bin_accr26}. On the other hand, supercritical accretion may be intermittent in nature, which limits the periods of particle accelerator activity. We shall discuss possible observational constraints of the intermittent activity of the extended jet of V4641~Sgr.

Below, we apply the minimal model with diffusive shock acceleration (DSA) of VHE particles to the supercritical black hole X-ray transient V4641~Sgr and discuss the perspective to test the model against future high angular resolution sensitive X-ray observations. In section \S \ref{Xrays} we present and analyze new data from the SRG/eROSITA all-sky survey in the 2–6 keV energy range together with archival hard X-ray fluxes detected by INTEGRAL and Swift BAT from V4641~Sgr vicinity. The data are used together with the radio emission and VHE gamma-ray detections to constrain the models of CR acceleration in the source. Simulations of the structure of the MHD flow produced by the interaction of a transrelativistic conical plasma jet with an isotropic wind of the accretion disk, which formed strong shocks at a distance of tens of parsecs from the black hole, are presented in  \S \ref{MHDs}. 
Simulations of the spectra of particles and turbulent magnetic fields amplified by cosmic ray-driven instabilities by the nonlinear DSA mechanism at strong recollimation shocks are presented in \S \ref{Kinetic}. Modeling of broadband non-thermal emission produced by VHE particles accelerated in-situ by the shocks in the extended jets and wind is discussed in \S \ref{rad}. A possible role of extended outflows of the stellar mass black holes accreting at super-Eddington rate as sources of VHE cosmic rays is discussed in \S \ref{CRs} and \ref{Sum}.

\section{X-ray observations}\label{Xrays}

\begin{figure*}
        \includegraphics[clip=true,width=\textwidth] 
        {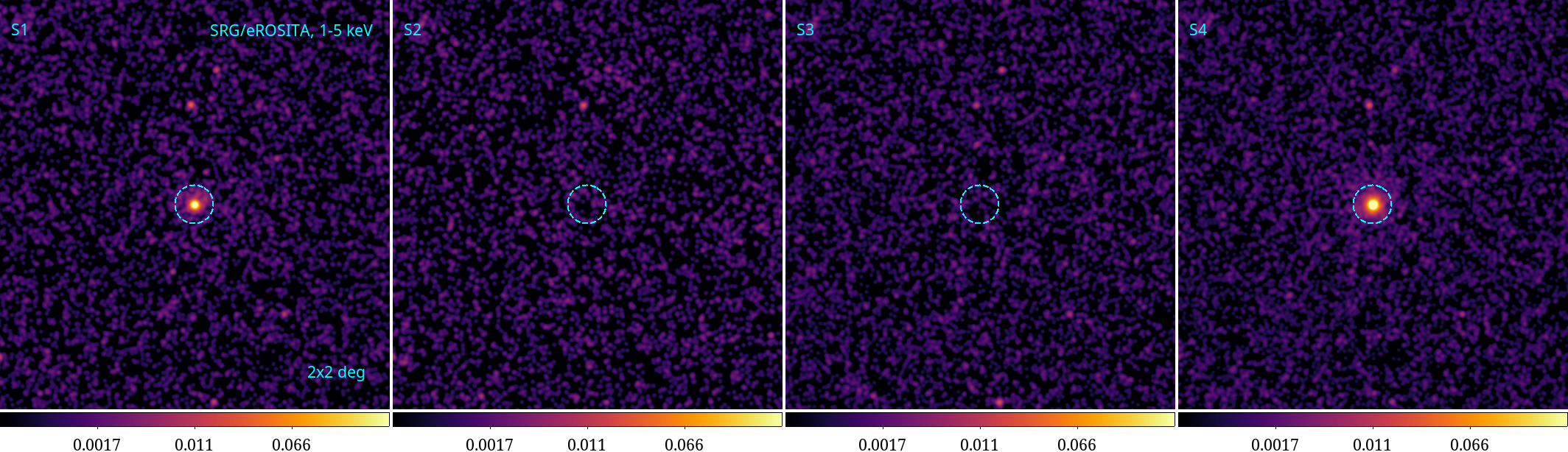}
\caption{X-ray images in the 1-5 keV band of the V4641~Sgr field ($2^{\circ}\times$$2^{\circ}$, Galactic coordinates) obtained during the 4 scans separated by half a year in the SRG/eROSITA all-sky survey. 
   The position of V4641~Sgr is shown with a circle ($6'$ radius) centered at the source. The compact source was bright during the first and the fourth scans, and absent in the second and third scans. The data of the second and the third scans are used to put constraints on diffuse {X-ray} emission from the TeV lobes.}
   \label{erass}
\end{figure*}

Observations of the microquasar V4641~Sgr with XRISM in 2024 revealed, in addition to the bright central source, a tentative hint of extended X-ray emission  \citep{2025ApJ...978L..20S}.
The integrated flux of this emission (approximated with a Gaussian with $\sigma=10'$) is $(4–6) \times 10^{-12} \enf$ in the 2–10 keV range \citep[with the source region of surface area 629.4 arcmin$^2$, ][]{2025ApJ...978L..20S}. However, this extended component is subdominant to the emission components of the sky and particle emission background, and might be sensitive to the exact modeling of the PSF shape \citep{2025ApJ...978L..20S}. Therefore, to better constrain the extended emission, sensitive high angular resolution observations covering the region of interest are required, ideally at an epoch when the central source is in the low-flux state.

\begin{figure}
\includegraphics[width=0.49 \textwidth]{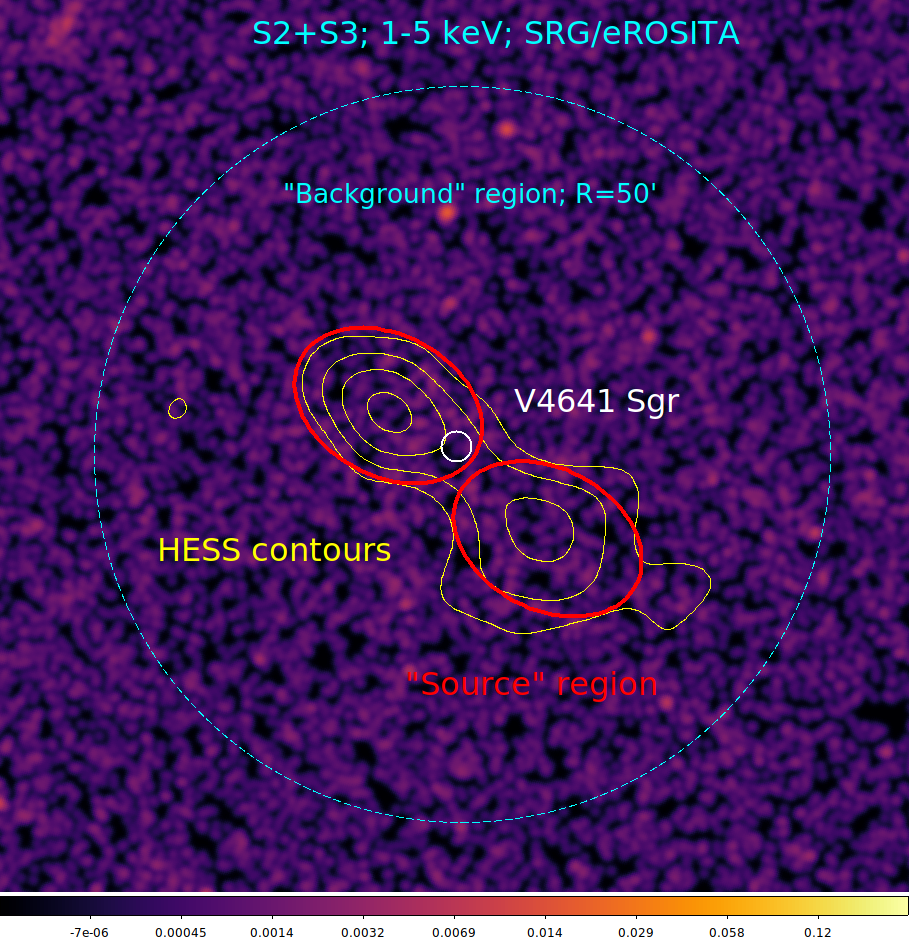} 
  \caption{\label{eR_map}
 The eROSITA 1-5 keV map for the two scans S2 and S3, when the central black hole was dim, enabling a search for extended faint X-ray emission from the regions where VHE radiation was detected. The yellow contours show the distribution of the VHE emission from the H.E.S.S. observations (significance levels $S=$3, 5, 7, and 9) reported in \citet{V4641_HESS_2026A&A...706A...8A}. The red ellipses show the regions used to extract X-ray flux. The large circle depicts the region used to estimate the X-ray background (excluding the elliptical regions).}
 \end{figure}

To constrain the diffuse X-ray emission from a wide field around V4641~Sgr, we used the SRG/eROSITA \citep{2021A&A...656A.132S,2021A&A...647A...1P}  data accumulated during the all-sky survey. The source was observed 4 times in individual surveys separated by half a year. The central source V4641~Sgr demonstrated violent epoch-to-epoch variability, being bright during the scans S1 and S4 and dim during the scans S2 and S3 (Fig.~\ref{erass}). In the context of the present study, we are interested in detecting or constraining a long-lived extended non-thermal X-ray emission away from the central source, which might accompany the detected VHE emission, as is the case in the microquasar W50/SS~433. The region of interest is defined by the flux contours of the VHE emission detected by the H.E.S.S. observatory \citep{V4641_HESS_2026A&A...706A...8A} shown in Fig.\ref{eR_map}. Given the shape of the extended part of the eROSITA telescope point spread function, the choice of the dim periods of the central source allows us to reduce its contribution to the putative diffuse emission. We extracted the spectrum from the regions shown in Fig.\ref{eR_map} and used a large annulus ($r=10’-30’$) as the background. 

The observed flux in the source region was found to be consistent with that in the background. By assuming a power law with the photon index of 2, we set a $3\,\sigma$ upper limit on the excess flux in the 2-6~keV band of $3.7 \times 10^{-12}\,\enf$ ($5.4 \times 10^{-12}\,\enf$ in 2-10 keV). Above 2~keV, the impact of the low-energy photoelectric absorption towards V4641~Sgr on the observed flux is subdominant, and no extra corrections are needed. This upper limit almost coincides with the flux claimed for the XRISM detection. Given the better angular resolution of eROSITA and the absence of the bright central source in the two used scan observations, we use the limit obtained below to constrain the X-ray flux from the extended PeV source.

\begin{figure}
    \begin{minipage}{.49\textwidth}
        \includegraphics[width=\textwidth,clip=true,trim=2cm 7.8cm 4cm 6.cm]{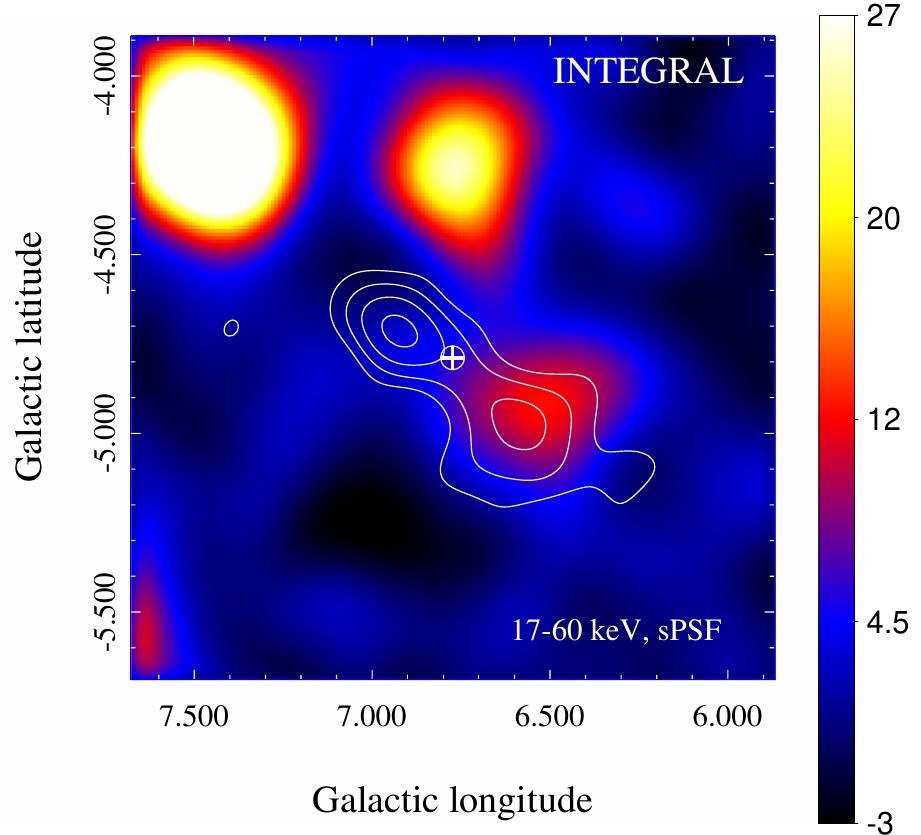} 
   \end{minipage}
   \caption{The image of statistical significance in the 17-60 keV energy range, smoothed with INTEGRAL/IBIS's PSF constructed from 17 years of hard X-ray all-sky survey   \citep{2022MNRAS.510.4796K}. The circle with a cross in the center shows the location of V4641~Sgr, while the contours show the H.E.S.S. significance levels ($S=3,5,7$ and $9$) from \citep{V4641_HESS_2026A&A...706A...8A}.The source IGR~J18193-2542 in the vicinity of V4641~Sgr is detected with the 17-60 keV flux $F_{17-60}=6.4\cdot10^{-12}$ erg/s/cm$^2$ and signal-to-noise ratio $S/N=12.0$, as 
   listed in the corresponding catalog \citep{2022MNRAS.510.4796K}.}
   \label{INT}
\end{figure}

Interestingly, there is a source IGR~J18193-2542 detected by the INTEGRAL soft gamma-ray imager \citep[ISGRI,][]{2003A&A...411L.141L} above 17 keV and listed in the second IBIS/ISGRI  catalog by \citet{2006ApJ...636..765B}. It has a flux of $(0.64 \pm 0.05) \times 10^{-11} \enf$ in the range 17 - 60 keV reported in the INTEGRAL/IBIS 17-yr hard X-ray all-sky survey by \citet{2022MNRAS.510.4796K}. The image of IGR J18193-2542 and the position of the black hole in V4641~Sgr are shown in Fig. \ref{INT}. 
Similarly, there is a source, Swift~BAT J1818.7-2553, with a flux of 6.23 (0.25, 12.26) $\times 10^{-12} \enf$ in the energy range 14-195 keV with the photon index 1.54 reported after 105 months of observations  \footnote{https://swift.gsfc.nasa.gov/results/bs105mon/1534}. Chandra did not detect a point source within the $1.5\arcmin$ INTEGRAL IBIS/ISGRI error circle of IGR J18193-2542 with an upper limit of $1.4\times 10^{-14} \enf$ in the 0.3-10 keV photon energy range \citep{2008ApJ...685.1143T}.

The luminosity of each of the extended X-ray jets in SS433/W50 is at the level of few $\sim10^{34}$ erg~s$^{-1}$ \citep{Brinkmann2007,Safi-Harb2022,2026A&A...707A.278S}, which would correspond to the flux of $\sim5\cdot10^{-12}$ erg~s$^{-1}$~cm$^{-2}$ at the assumed distance to V4641~Sgr of 6 kpc. This is significantly above the conservative upper limit set by SRG/eROSITA, implying that the unified model should be capable of explaining possible variations in the TeV-to-X-ray ratio. On the other hand, the detection of the hard X-ray source with INTEGRAL IBIS/ISGRI and Swift BAT in the vicinity of V4641~Sgr, if confirmed, might be an indication of a relatively harder spectrum of the emitting particles.
In what follows, we use the eROSITA, Swift BAT, and INTEGRAL fluxes and upper limits to constrain the model of the V4641~Sgr spatially-extended X-ray emission and to study possible intermittent character of particle acceleration in the microquasars accreting at the supercritical rate. 


\section{MHD model of extended jets and winds}\label{MHDs}
\begin{figure}[h]
\includegraphics[scale=0.33]{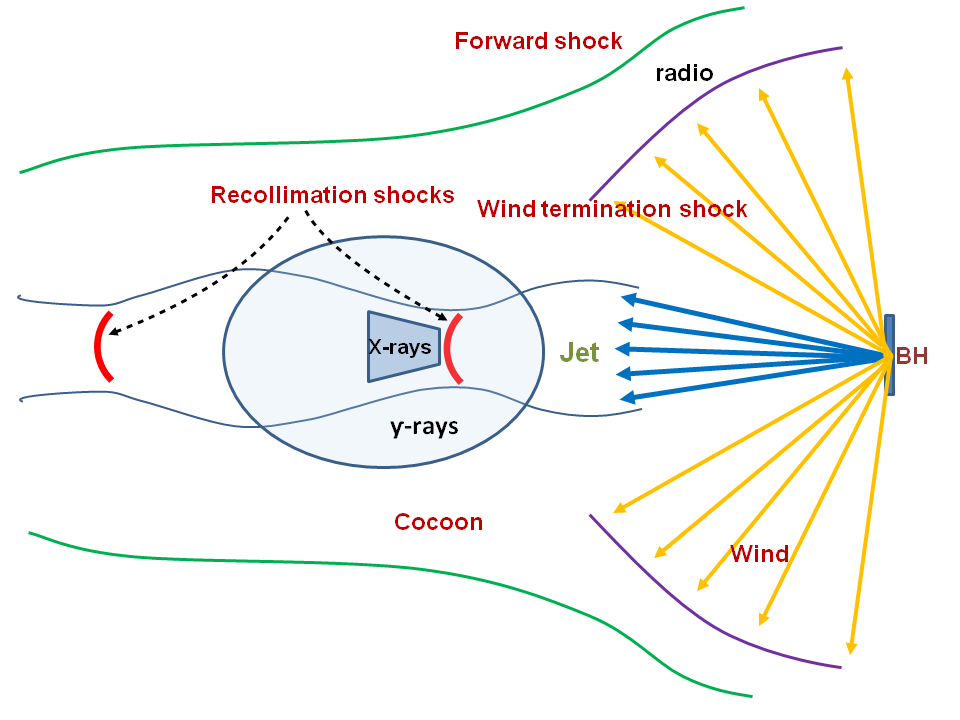}
\caption{Schematic representation of the structure of extended energetic plasma outflows producing the non-thermal radiation of the nebula around the microquasar V4641~Sgr. The source of energy and momentum - a stellar-mass black hole in the binary system accreting at a super-Eddington rate is labelled as BH. The fast and narrow collimated transrelativistic jet from the polar outflow is interacting with the quasi-isotropic (slow wind)  outflow of velocity $\sim 3,000 \kms$ from the black hole accretion disk. The wind termination and forward shocks, as well as the jet recollimation shocks, are shown. The cone area in the diagram near the shock corresponds to regions of synchrotron X-ray radiation of the VHE electrons in the turbulent magnetic fields amplified by the instabilities of cosmic rays accelerated at the strong shocks. The regions of VHE gamma-ray radiation by the inverse Compton scattering of PeV regime electrons are shown in the sketch.
} \label{Scheme}
\end{figure}

To study relativistic particle acceleration and radiation, we simulated the structure of the interaction of the extended outflows from the supercritically accreting black hole binary with interstellar matter following the model described in \citet{Churazov, SS433_min_model_2025PhRvD.112f3017B} using the numerical MHD code PLUTO \citep{Mignone2007}. We assume that the microquasar ejects matter in the form of a slow wind and a fast collimated jet. The slow wind has a wide solid angle, while the fast jet occupies a smaller solid angle. A schematic representation of the system is shown in Fig.~\ref{Scheme}.

\begin{figure}[t]
\includegraphics[scale=0.5]{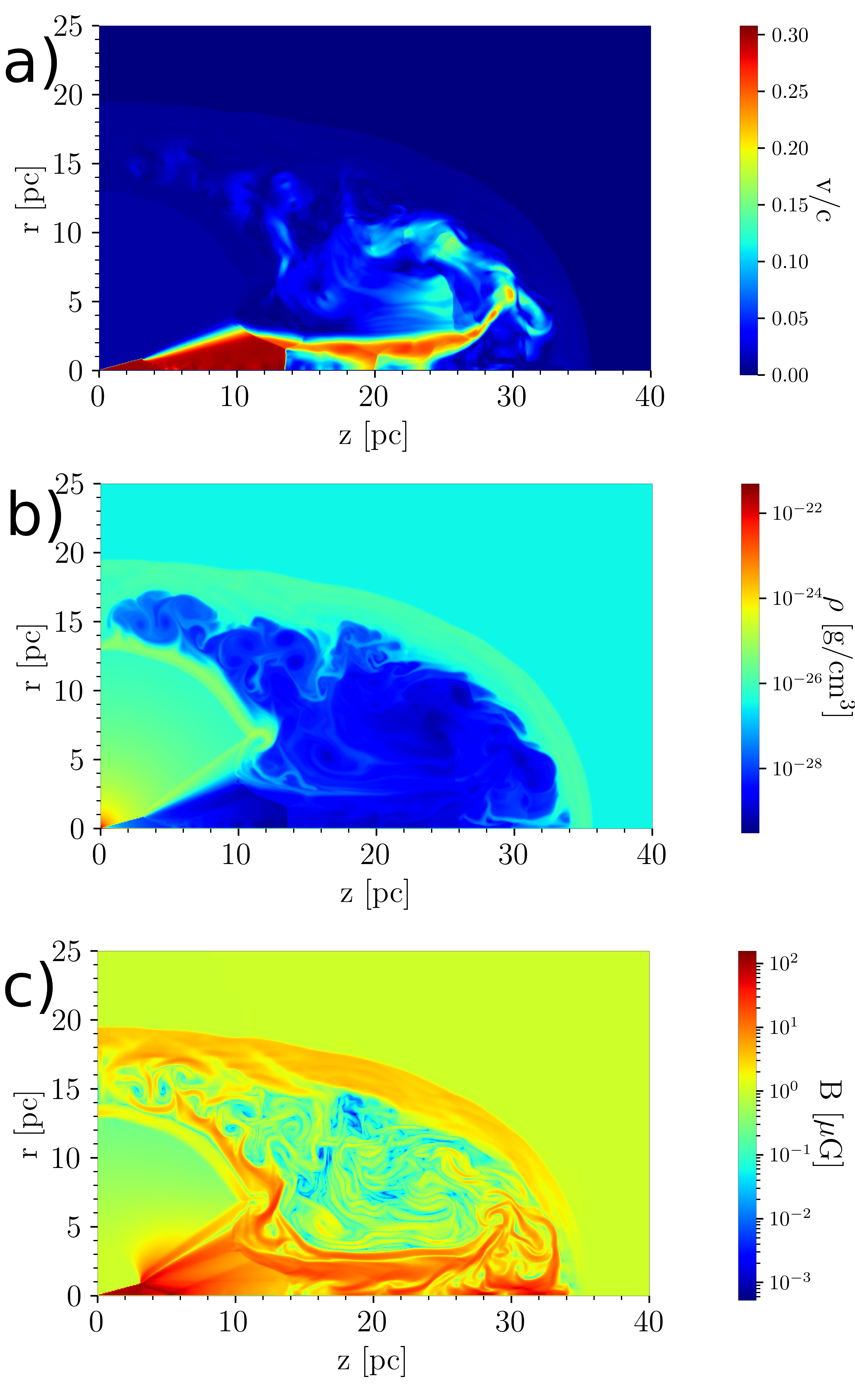}
\caption{Magneto-hydrodynamic maps of the flow from the axisymmetric simulations of jet- wind interactions. The jet velocity was initiated with a velocity of 0.3c and the power of $2\cdot10^{39}~ \ergs$. The wind velocity was initiated at 3000 $\kms$. Panel a -  map of the flow velocity, panel b - map of density, panel c - map of the magnetic field in the flow. } 
\label{MHD}
\end{figure}

The parsec-scale extended flow morphology was simulated in a two-dimensional setup in cylindrical coordinates with 2000 grid points along the z-axis and 1000 along the radial axis. Given the observed position of V4641~Sgr, its surrounding interstellar medium is set to be homogeneous with number density $n_0=2\cdot10^{-3}~\cmc$, temperature $T_0=10^6~K$, and magnetic field $B_0=10^{-6}$~G. Internal boundary conditions to model the extended fast jet and wind are set on the sphere with radius $1\cdot10^{19}$~cm, at which the velocity, density, and pressure are held constant. The kinetic powers of both the slow wind and the sum of the powers of the two extended jets are set to $2\cdot10^{39}~ \ergs$, and the velocity of the jet is $0.3~c$, where $c$ is the speed of light.
For polar angles closer to the z-axis than the jet half-angle $\theta_{jet}=15^{\circ}$, parameters at the boundary sphere correspond to the jet, and for larger angles, to the slow wind from the accretion disk.

Multiwavelength observations of the outflow from the outer parts of the accretion disks in the close vicinity of the transient accreting X-ray source V4641~Sgr, together with the other galactic black hole binaries  V404~Cyg and GRS~1915+105, demonstrated lines with strong P-Cyg profile during different observed epochs with outflow velocities $\sim$ 3000 $\kms$ \citep{accr_disc_winds_SSRv26,V4641_wind_XRISM_2025arXiv250817541P}.
Therefore, in our MHD and kinetic simulations, we used the velocity of the slow wind equal to $3000~\kms$. The results of the MHD simulations are presented in Figure~\ref{MHD}. 

\begin{figure}[h]
\includegraphics[scale=0.6]{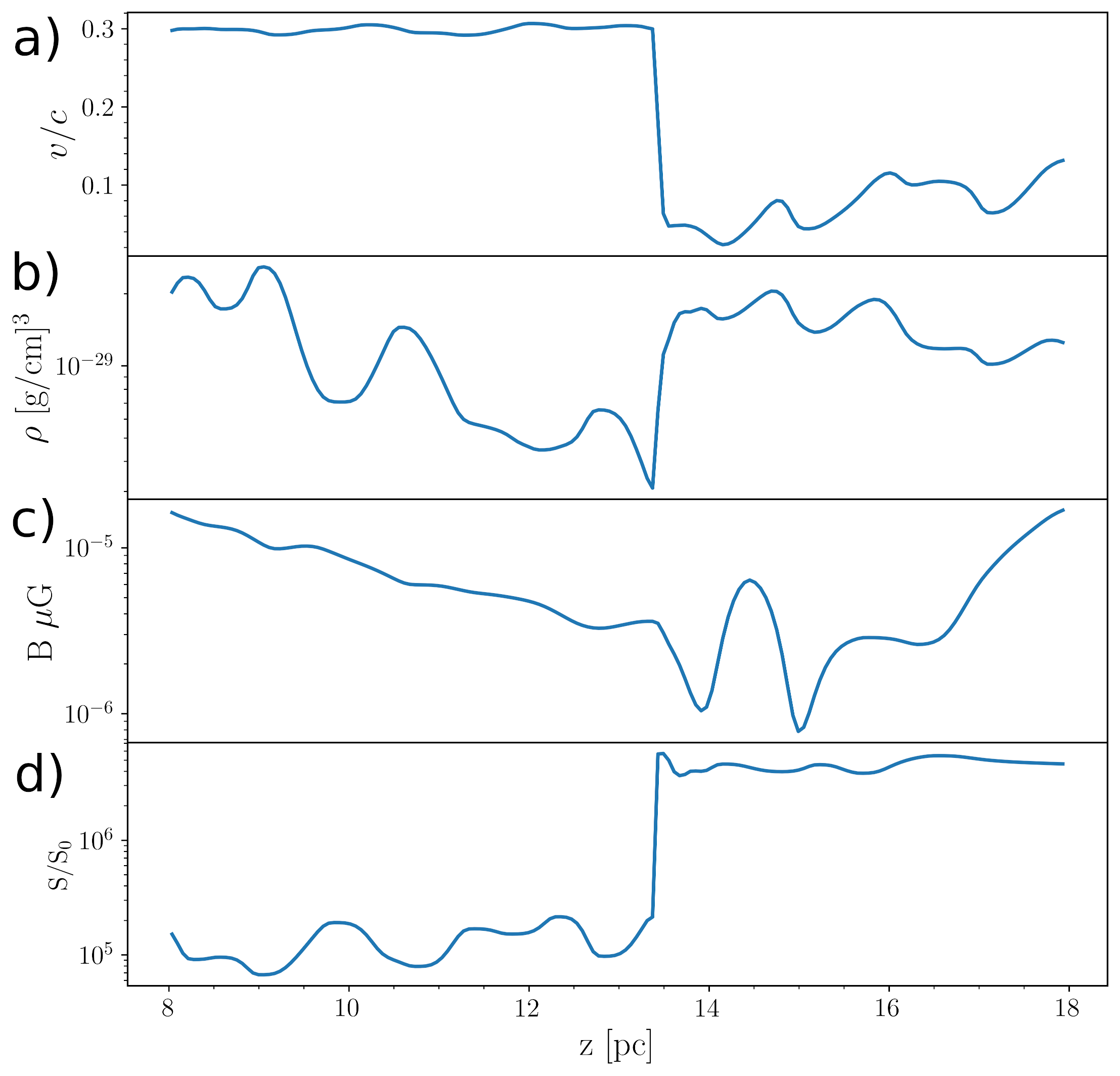}
\caption{The flow profiles along the z-axis around the jet termination shock. Panel a - velocity, panel b - density, panel c - magnetic field, panel d - entropy} 
\label{profile_window}
\end{figure}

\begin{figure}[h]
\includegraphics[scale=0.6]{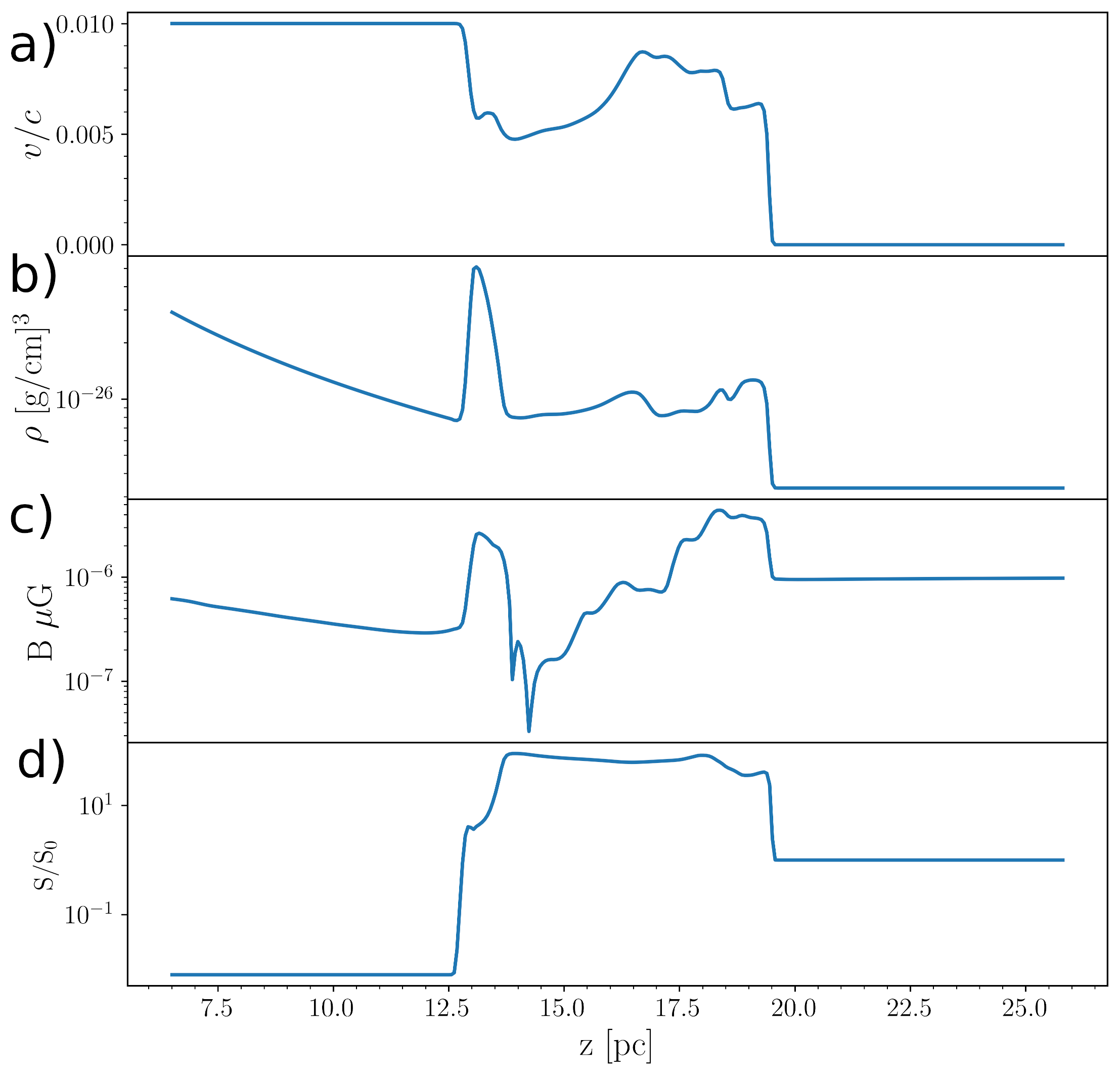}
\caption{The flow profiles along the r-axis (transverse to z) with the broad wind termination and forward shocks. Panel a - velocity, panel b - density, panel c - magnetic field, panel d - entropy} 
\label{profile_r_window}
\end{figure}

The simulated profile of the flow (from Fig.~\ref{MHD}) along the jet axis with recollimation and termination shocks is shown in Fig.~\ref{profile_window}, while the profile along the axis transverse to the jet, featuring the wind's termination and forward shocks, is shown in Fig. \ref{profile_r_window}. Simulations show the presence of wind termination and forward shocks at distances, corresponding to radio emission observed by MeerKAT \citep{MeerKaT2026A&A...706A.283G}, and prolong structures created by jets with recollimation shocks, which might be associated with the emission detected by HAWC \citep{V4641_HAWC_2024Natur.634..557A}, LHAASO \citep{LHAASO_MQs_2025NSRev..12af496L} and H.E.S.S. \citep{V4641_HESS_2026A&A...706A...8A}.
 Using these figures, one can identify the strong shocks which can accelerate ultra-relativistic particles to produce the gamma-ray-bright regions in the jet and the radio-bright regions in the spherical wind. 

In Fig. \ref{profile_window}, panel (d), the jet's strong recollimation shock wave is clearly visible at a distance of about 14 pc from the source as an entropy jump. The termination shock of the slow wind is clearly visible near 12.5 pc in Fig.~\ref{profile_r_window}, panel (d), and the forward shock of the slow wind near 19.5 pc. Both are marked by the entropy jumps. Using the profiles from Figs. \ref{profile_window} and \ref{profile_r_window}, we determine the velocities, densities, and magnetic field strengths upstream of the shock waves. The radial profiles of the velocity, density, and magnetic field upstream of the shock are used below (see \S\ref{Kinetic}) in the Monte-Carlo modeling of particle acceleration at the shocks and magnetic field amplification by cosmic ray-driven instabilities.

The extended slow wind interacts with the surrounding interstellar medium 
forming a forward and wind termination shocks with a region in between that separates the wind and the shocked ambient gas. 
For the wind and ambient medium parameters within the minimal model described above, the speeds of the termination and forward shocks were obtained to be 
several thousands $\kms$ (see Fig.\ref{profile_r_window}). 
These extended strong shock waves are fast enough to accelerate particles 
to several tens of TeV by the Fermi DSA mechanism \citep{BE87,JE91,Bell04}. Electrons accelerated by these shock waves emit a broad spectrum of synchrotron radiation, which matches the observed radio emission and extends to X-ray energies under some parameter range to be tested 
by sensitive X-ray observations. The inverse Compton radiation produced by these electrons would be in the GeV-TeV regime.

The fast and narrow collimated jet in the model has a velocity of 0.3c. A strong normal recollimation shock wave of the fast jet occurs near the position of the termination shock of the extended slow wind. This shock wave accelerates particles of the PeV regime energies. Electrons accelerated to such energies emit photons with energies of hundreds of TeV up 
to PeV due to the IC mechanism, while their synchrotron radiation is in the X-ray range.

The presence of significant X-ray variability (see Fig. \ref{eR_map}) and strong flares in V4641~Sgr is a distinct observed feature compared to W50/SS~433. The integrated effect of the flares may produce substantial time variability of the global system outflows at parsec distances. The variable plasma velocity and density in the collimated jet will cause restructuring of the recollimation shocks, which in some cases leads to the disappearance of a strong normal shock.

The normal shock with a strong jump of the entropy is the site of PeV particle acceleration by the Fermi DSA  process with turbulent magnetic field amplification. The life-time of leptons of energy ${\rm E}$ in the jet and surrounding cocoon $\sim$ 1,000~ ${\rm (E/PeV)^{-1}}$ ${\rm[f(E/PeV) +(B/3.3\mu G)^2]^{-1}}$~ yrs, where $\rm f(E/PeV)$ is the normalized effective energy density of the ambient radiation field corrected for the Klein-Nishina effect with $\rm f(1) \approx  0.03$,  $\rm f(0.1) \approx  0.3$ and $\rm f  \approx  1$ for $\rm E \leq 10$ TeV (if we consider only CMB in the photon energy density). It allows for both synchrotron X-rays and gamma-rays from the inverse Compton scatterings over a time period longer than 1 kyr in the cocoon with a low magnetic field and of a few hundred years inside the jet after the DSA acceleration ceased. The synchrotron radiation will be fading not only because of particle energy losses but also because of the decay of the turbulent magnetic fields amplified by cosmic ray-driven instabilities during the DSA process.

\section{Kinetic simulation of particle acceleration}\label{Kinetic}

 To study a possible parameter space and to confront with the available observations, we model two somewhat different jets -  jet A and jet B with slightly different parameters. Jet B has a higher density and magnetic field; thus it produces more X-ray synchrotron radiation, while jet A produces more inverse Compton gamma radiation. The differences can be explained by the fact that the recollimation shocks in jet A and jet B are located at different distances from the central source, which is supported by the asymmetry of the system observed by H.E.S.S. \citep{V4641_HESS_2026A&A...706A...8A}. 
 The electrons that are accelerated by the shock waves produced by fast outflows radiate mainly from the downstream of these shock waves. In this section, we calculate the spectra of accelerated ions and electrons directly at the shock fronts using a Monte Carlo code at parameters near the shocks obtained in the MHD simulation described in the previous section. Then the morphology of radiation and the spectra are calculated using the spectra of accelerated electrons and ions obtained in this section.

\begin{figure}[h]
\includegraphics[scale=0.3]{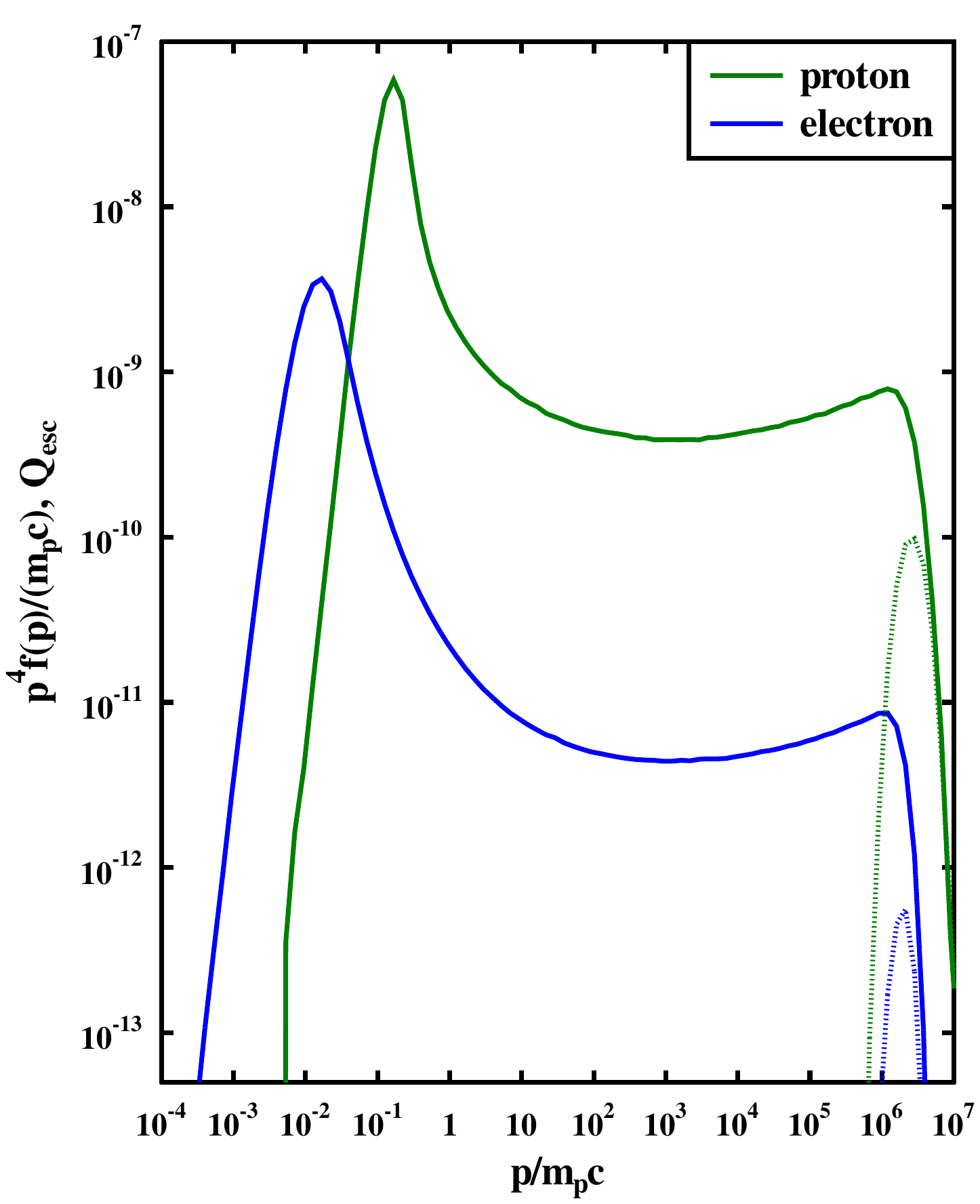}
\caption{The distribution functions of electrons and protons accelerated at the strong recollimation shock front of the extended transrelativistic jet A of velocity 0.3c (see Figs.\ref{MHD}, \ref{profile_window}) are shown as solid curves. The spectra for Jet B have the same shape in the VHE range but differ in amplitude. The spectra were simulated with the non-linear Monte Carlo DSA model. The fluxes of escape particles through the FEB are shown by dashed curves (\ref{Qesc}).} 
\label{pdf_jet}
\end{figure}

\begin{figure}[h]
\includegraphics[scale=0.3]{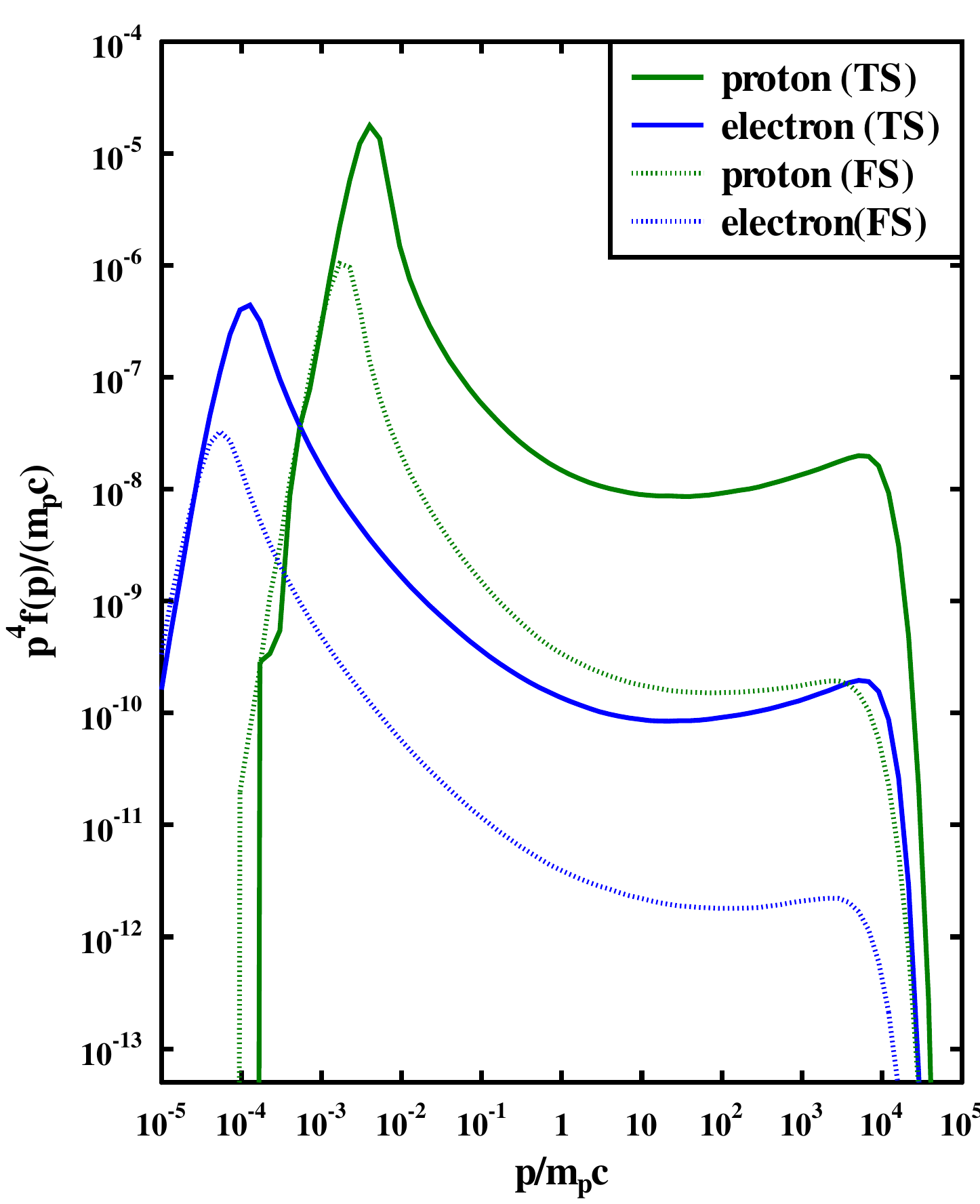}
\caption{The distribution functions of electrons and protons accelerated at the termination shock front of the broad wind of velocity 3,000 $\kms$ (see Fig.\ref{profile_r_window}) simulated with a non-linear DSA model are shown by solid curves. The distribution functions of electrons and protons accelerated at the forward shock driven by the wind are shown by dotted curves.} 
\label{pdf_SW}
\end{figure}

Particle acceleration and magnetic field amplification by shocks are calculated with a non-linear Monte Carlo model \citep[][]{Bykov3inst2014,Bykov2022Univ}. A distinctive feature of the model is the kinetic calculation of the full ion distribution function, accounting for nonlinear feedback effects on the shock wave structure and magnetic field amplification. This model iteratively determines a self-consistent plane-parallel, quasi-steady shock flow solution, the spectra of particles, and the amplified magnetic field turbulence. The magnetic field amplification in the model is due to cosmic ray-driven instabilities produced by the current anisotropy of accelerated particles.  In the code, the laws of conservation of energy and momentum flows near the shock are achieved in an iterative process. In the Monte Carlo model, the particle free escape boundary (FEB) is located far upstream of the shock. The distance from the FEB to the shock front is denoted by $L_{FEB}$. In the strong transrelativistic shocks in the electron-ion plasma, the electrons have little feedback effect on the shock flow. Electron acceleration is considered in the test particle regime.  

In our simulation of electron spectra, we assumed that their injection rates are in the range 0.001- 0.01 (by particle numbers) relative to protons, consistent with the available estimations from particle-in-cell simulations of transrelativistic shocks \citep[see e.g.][]{2019MNRAS.485.5105C,Bykov2022Univ,transrel_sironi_2026ApJ...998..149J}. The energy losses of accelerated relativistic electrons due to inverse Compton and synchrotron radiation are taken into account. 

When modeling particle acceleration by the jet recollimation shock, the following parameters were selected near the FEB ($L_{FEB}=8$ pc): the background proton number density is $\displaystyle  10^{-6}$ cm$^{-3}$ (jet A) and $\displaystyle  4\cdot 10^{-6}$ cm$^{-3}$ (jet B). The rms magnitude of the turbulent magnetic field in the jet is 1 $\mu$G, and the regular magnetic field is 0.1 $\mu$G. When the regular magnetic field is less than the turbulent one, its value has little effect on the calculation results, which is required due to the technical design of the code. The background jet plasma velocity relative to the shock front in the far upstream is 0.3c. The turbulent magnetic field has a Kolmogorov spectrum at the FEB with an energy-carrying scale of about 5 pc. 

The simulated proton and electron spectra are shown in Fig. \ref{pdf_jet}. The flux of accelerated particles escaping through the FEB in the far upstream is 
\begin{equation}
\label{Qesc}
    Q_{esc}=\frac{j_{p}^{cr}(L_{FEB},p)p^{4}}{4\pi m_{p}c^{2}},
\end{equation}
\begin{equation}
\label{jcr}
   j^{cr}(z)=\int_{0}^{\infty} j_{p}^{cr}(z,p)p^{2}dp,
\end{equation}
where $p$ is particle momentum, $m_{p}$ is the proton mass, and $j^{cr}(z)$ is the current of the accelerated particles at a given point. The particle distribution functions are normalized as follows
\begin{equation}
\label{fcrnorm}
   n(z)=4\pi\int_{0}^{\infty} f(z,p)p^{2}dp,
\end{equation}
where $n(z)$ is the number density of accelerated particles.

The kinetic description of particle transport in the iterative shock structure solution allows one to estimate the acceleration time of electrons and protons up to the highest energy of about a few PeV, which is limited here by the jet transverse size. The acceleration time is about 300 years. 
The magnitude of the downstream fluctuating magnetic field amplified by cosmic ray driven instabilities is 10-20 $\mu$G for different jet parameters. The turbulent magnetic field downstream of the strong shock is localized in the shock front vicinity and decreases with the distance from the shock as seen in the bottom panel in Fig. \ref{profile}.     
When the shock acceleration process ends, e.g., due to jet parameter variation, the turbulent field will decay within about 300 years, i.e., much faster than the lifetime of accelerated electrons in the cocoon.   

We also simulated particle acceleration by the diffusive shock acceleration mechanism at the forward and termination shocks of the slow isotropic wind extended to tens of pc from the accretion disk. A termination shock of the supersonic wind is formed inside the large-scale cocoon where the magnetic fields are below the $\mu$G range, as indicated by MHD simulations presented in Fig. \ref{MHD}. In the acceleration model for both shocks, $L_{FEB}=1$ pc, the rms of the turbulent magnetic field is 1 $\mu$G for the forward shock and 0.2 $\mu$G for the termination shock, and the homogeneous component of the magnetic field is about 0.2 $\mu$G. The turbulent magnetic field has a Kolmogorov spectrum on the FEB with an energy-carrying scale of about 1 pc. The background plasma velocity relative to the shock front in the far upstream is 3000 km/s for the termination shock and 1000 km/s for the forward shock. The background proton number density in the far upstream is $\displaystyle  10^{-2}$ cm$^{-3}$ for the termination shock and $\displaystyle  2\cdot 10^{-3}$ cm$^{-3}$ for the forward shock. The
protons and electron spectra obtained as a result of the simulation are shown in Fig. \ref{pdf_SW}.

\section{Non-thermal emission of the extended jets and the wind}\label{rad}

To model the emission spectrum from the source, we use the particle spectrum at the shock front obtained with a non-linear Monte Carlo DSA model with magnetic field amplification by cosmic ray-driven instabilities. The simulated electron and proton spectra at the shock are shown in Fig.~\ref{pdf_jet} and \ref{pdf_SW}.
The emission model accounts for particle energy losses due to synchrotron and inverse Compton radiation for particles transported to the shock downstream during the time of source activity $t_{\rm sr}$, which is determined by the lifetime of the strong shock. The spectral shape  of leptonic emission depends on the relation between $t_{\rm sr}$ and the time of particle energy losses.  
 
To evaluate the particle distribution function at different points downstream, we neglect diffusion (advection dominates on scales larger than several parsecs) and assume that the process is stationary. With these assumptions, we can easily evaluate how the energy of each particle changes during its advection downstream along the z-axis.

\begin{equation}
    E(z) = \frac{E\left(0\right)}{1+E\left(0\right)k(z)},
\end{equation}

where $\displaystyle k(z) = \int_0^z \frac{4\sigma_T}{3m_e^2c^3}\left(B(z)^2/8\pi + U_{ph}\right) \frac{dx}{v(z)}$. 

Thus, using the conservation of number of particles, we can express the electron distribution at a distance $z$ from the shock in the downstream as

\begin{equation}
    f(z,E)=f\left(0,\frac{E}{1-k(z)E}\right)\frac{1}{\left(1-k(z)E\right)^2},
\end{equation}

where $f(0,E)$ is the distribution function of electrons at the shock front. Knowing electrons distribution function, we can integrate synchrotron and inverse Compton radiation through the volume of the downstream region and obtain the full emissivity of the source.

\subsection{Very high energy emission of the extended jets in V4641~Sgr}

\begin{figure}[h]
\includegraphics[scale=0.6]{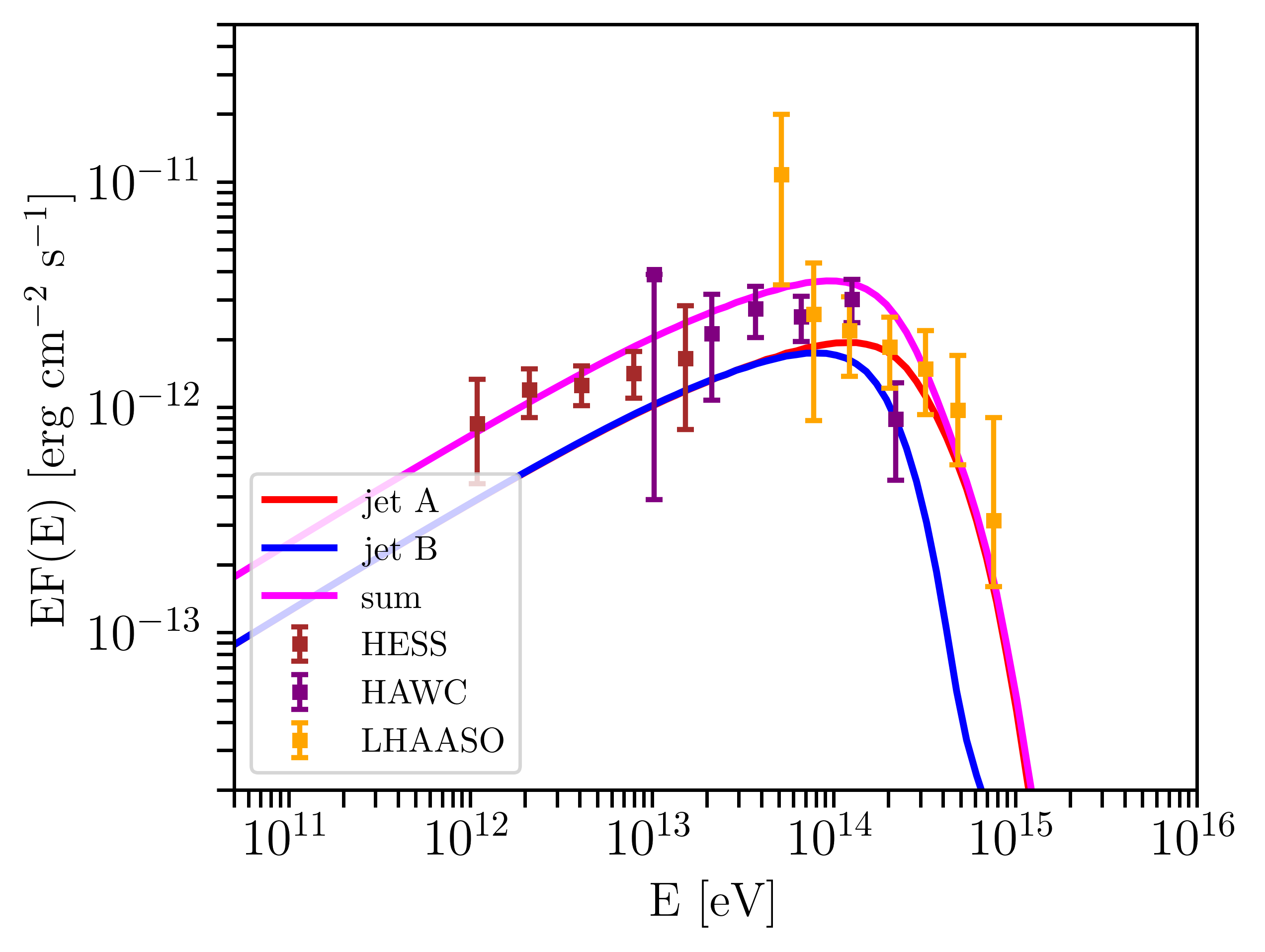}
\caption{Simulated high energy spectrum of the inverse Compton radiation from VHE electrons accelerated (their spectra are shown in Fig.\ref{pdf_jet}) in the shocked jet of velocity $0.3\,c$ shown with solid lines together with gamma-ray fluxes measured  by H.E.S.S. and LHAASO observatories in V4641~Sgr.} 
\label{compton}
\end{figure}

The non-linear diffusive shock acceleration mechanism discussed above simultaneously predicts both the spectra of ultra-relativistic particles and the associated magnetic turbulence produced by cosmic ray-driven instabilities in the shock upstream. The energy density of the magnetic turbulence just behind the shock front in the nonlinear DSA models is about a few percent of the shock's ram pressure. In the jet model of  V4641~Sgr, magnetic turbulence of magnitude 10-20~$\mu$G is produced at the strong relollimation shock, and the maximal energies of the accelerated particles $\gsim$ 1 PeV are achieved. The accelerated leptons can emit very high energy photons via inverse Compton scattering. However, if the magnetic field approaches the magnitude $\gsim 20~\mu$G, then the synchrotron losses dominate and the fraction of energy emitted as PeV photons decreases. To evaluate the emissivity of inverse Compton radiation at energy $E$, we use the equations for isotropic electron and photon distributions, derived by \citet{Jones68} for the isotropic distributions of accelerated electrons and the background photons

\begin{equation}
    \begin{split}
	\frac{dW(E)}{dt dE dV}=E\int \frac{2 \pi r_e^2 \beta_e c}{E_{ph} E_e^2} f_{ph}(E_{ph})f_e(E_e)\times\\
\left(2 q~ \ln(q)+1+q-2q^2+\frac{q^2(1-q)\Gamma^2}{2(1+q\Gamma)}\right)dE_{ph} dE_e dV,
    \end{split}
\end{equation}

where $E_{ph}$ is seed photon energy, $E_e$ - seed electron energy, $\gamma_e$ - seed electron Lorentz-factor, $\Gamma=4E_{ph}\gamma_e/m_e c^2$, and $q=E/((\gamma_e m_e c^2-E)\Gamma)$. In Figs. \ref{synchrotoron} and \ref{compton} we present the spectra of synchrotron and inverse Compton radiation for two jet shocks which have slightly different magnitudes of the amplified magnetic fields in the downstream. The magnitude of the turbulent magnetic field in the shock downstream depends mainly on the jet plasma density in the far upstream, which may be variable in the jets. One model which we dubbed in Figure  \ref{compton} as jet A has $\sim 10~\mu$G producing most of high energy gamma rays, while another jet B with stronger magnetic field  $\sim 20~\mu$G produces higher fluxes of synchrotron radiation in Fig.~\ref{synchrotoron}  which may fit the INTEGRAL and Swift hard X-ray data if the sources IGR J18193-2542 and SWIFT BAT J1818.7-2553 are associated with the jet. The electron injection rates were the same in both cases. The keV X-ray fluxes derived in the model can certainly be detected and imaged with dedicated observations of current X-ray observatories, including polarimetric observations. The degree of polarization above 10\% can be expected in the bright X-ray knots associated with the jet recollimation shocks, as it was shown in \citet{SS433_min_model_2025PhRvD.112f3017B} for the X-ray knots in the W50/SS433 jet. Moreover, the predicted MeV radiation can be observed or constrained with the next generation of MeV detectors  \citep[see e.g.][]{2018JHEAp..19....1D}.

The very high energy gamma-rays from jet B with a magnetic field magnitude $\sim 20~\mu$G would have lower spectral cutoff energies below PeV (see the corresponding curve in Fig. \ref{compton}). The time of the source activity (i.e., the duration of shock acceleration of PeV regime particles) affects the spectral shape of gamma-rays in our model. To fit the observed spectral shape at energies below 100 TeV, which are in the thin target regime in our model, the source activity should not be longer than 3000 years.

We presented above the results of modeling of particle acceleration and non-thermal emission from a transrelativistic jet of velocity 0.3c interacting with a nearly isotropic wind. We also address the question of whether it is possible to fit the existing observations of the non-thermal emission of V4641~Sgr with another velocity of the transrelativistic jet, since the analysis of radio observations by \citet{Marti_V4641_VLA_jets26} revealed in the close vicinity of the accreting object the jets of estimated velocities about $(0.8 \pm 0.1)$c aligned with the direction to the gamma-ray sources. Namely, we also constructed models for jets of velocity 0.8c with the same kinetic power, interacting with the same isotropic non-relativistic wind outflow. The MHD modeling of a powerful weakly magnetized jet revealed the structure of the flow shown in Fig.~\ref{MHD} with a strong recollimation shock in the region where the jet interacts with the wind termination shock. Then we performed kinetic Monte Carlo simulations with the non-linear diffusive acceleration model, which again provided efficient acceleration of protons and electrons of PeV energies. The model reproduced the observed VHE gamma-ray radiation, as it is shown with a dashed curve in Fig. \ref{longspectrum}. Thus, the modeled transrelativistic jets of velocity $\gsim 0.3$c can explain the main observational data on the non-thermal emission from the V4641~Sgr nebula. 

\subsection{Synchrotron X-ray emission of transrelativistic jets}\label{Xrays1}
The PeV regime electrons accelerated by strong shocks in the fast jets radiate synchrotron X-ray photons in micro-Gauss range magnetic fields. This radiation was detected and studied in a few localized positions of the extended jets of the SS~433 microquasar \citep[e.g.][and the references therein]{Brinkmann2007,Safi-Harb2022,IXPE2024SS433,2026A&A...707A.278S}. Very high energy particle acceleration in strong fast shocks is accompanied by superadiabatic amplification of fluctuating magnetic fields by cosmic ray current-driven instabilities, as was simulated in the kinetic DSA model of SS~433 by  \citet{SS433_min_model_2025PhRvD.112f3017B}. In this model, the amplified fluctuating magnetic field is localized in the shock vicinity, which allows us to estimate the extension of the shock and jet opening angle by direct confrontation of the simulated synchrotron morphology against the X-ray and gamma-ray data.    

V4641 Sgr vicinity was observed by SRG/eROSITA and INTEGRAL \citep{2022MNRAS.510.4796K} and its flux (excluding the central source) of less than $\sim 4\times 10^{-12}\,\enf$ (3$\sigma$ upper limit) in the range 2-6 keV and does not exceed $(0.64 \pm 0.05) \times 10^{-11} \enf$ in the range 17 - 60 keV. The non-thermal synchrotron radiation component is determined by the spectra of accelerated leptons and the magnetic field structure.  

 In this paper, we assume that the magnetic field magnitude far upstream of the recollimation shock front is about $\mu$G, as it corresponds to relatively low magnetization ($\lsim 10^{-3}$) of the jet. Super-adiabatic amplification of turbulent magnetic fields due to cosmic ray-driven instabilities provided the amplitudes of the field directly behind the jet shock front of 10 $\mu$G for jet A and 20 $\mu$G  for jet B, respectively. The magnitudes of the cosmic ray-driven turbulence were obtained with the nonlinear Monte Carlo DSA model  \S\ref{Kinetic}. The amplified turbulent magnetic field then decays with distance from the shock front. 
 The turbulent field decay modeling for the magnetic field, similar to that in jet B, was described in detail by \citet[][]{SS433_min_model_2025PhRvD.112f3017B}. The characteristic decay length $L_{\ast}$ there was about 0.1 pc. 
 The simulated magnetic field profile provides the fluxes of hard X-ray radiation to fit the X-rays from INTEGRAL and Swift BAT sources  (assuming  their association with the jet B)  and is also consistent with the upper limits from the SRG/eROSITA telescope established in \S \ref{Xrays}

\begin{figure}[h]
\includegraphics[scale=0.6]{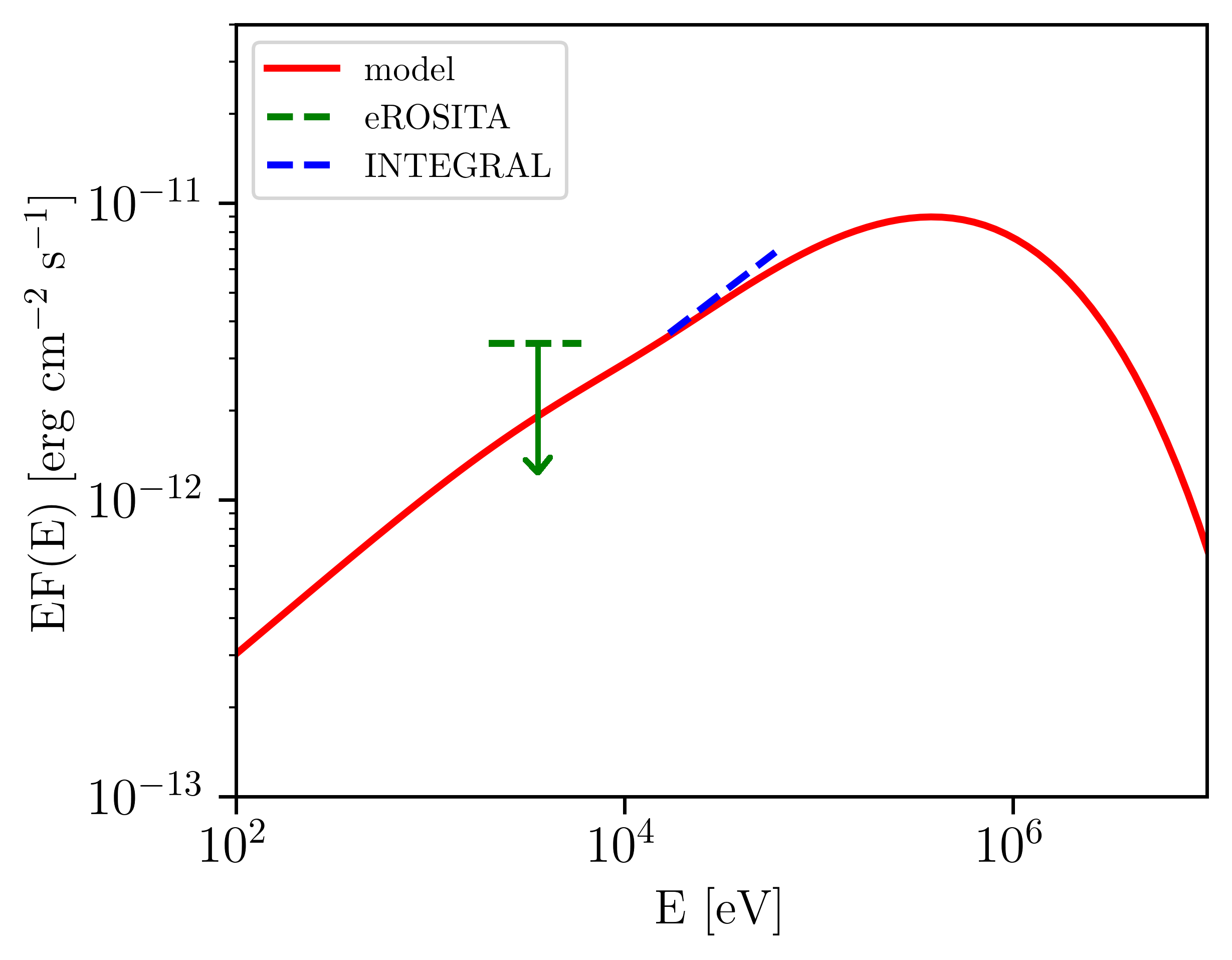}
\caption{Simulated spectral energy distribution of synchrotron X-ray radiation of electrons accelerated at the strong recollimation in the extended transrelativistic jet of velocity 0.3c in V4641~Sgr with the magnitude of the amplified turbulent magnetic field 20 $\mu$G (model jet B). This is the maximal flux for a persistent source consistent with the $3\sigma$ upper limit from SRG/eROSITA on the diffuse 2 - 6 keV emission from the extended region associated with the VHE radiation sources shown in Fig.~\ref{eR_map} and the fluxes of the hard X-ray sources  IGR J18193-2542  and Swift BAT J1818.7-2553  discussed in section \ref{Xrays}. The spectral fluxes of jet A with a lower magnetic field are somewhat below that of jet B (see Fig.~\ref{longspectrum}). The X-ray fluxes are expected to be lower than that shown here in the case of the intermittent evolution of the particle accelerator in V4641 Sgr as discussed in \S \ref{Xrays1}.} 
\label{synchrotoron}
\end{figure}

 Synchrotron emissivity (spectral density of energy radiated per unit time) per unit volume is evaluated using the standard equations \citep[see e.g.][]{1965ARA&A...3..297G}

\begin{equation} \label{SynchrotronEmission}
	\frac{dW(E)}{dt dE dV}=\int dE_e \frac {\sqrt {3}{e}^{3}n f_e(E_e) B \sin ( \theta)}{{m_e}{c}^{2}}
	\frac{E_e}{E_c}\int_{\frac {E_e}{E_c}}^{\infty }\it K_{5/3}(x)dx,
\end{equation}

where $\theta$ is angle between magnetic field and line of sight, $\displaystyle E_{c}$ is critical energy defined as $\displaystyle E_{c} = 3 h e^{2} B \sin(\theta) E^{2}/4\pi {m_{e}}^{3} c^{5}$, and~$K_{5/3}$ - Macdonald function.
 A comparison of the modeled synchrotron spectrum with the available observational data is shown in Fig. \ref{synchrotoron}. The synchrotron spectrum has a broad bump component due to the inhomogeneity of the magnetic field in the radiation zone. Note that the X-ray fluxes shown in 
Fig. \ref{synchrotoron} were derived assuming the jet kinetic power of 2$\times 10^{39} \ergs$. We show that in this case the model flux of synchrotron emission does not exceed the upper limit derived from the eROSITA/SRG survey analysis. The lower-power jet with the same velocity can still accelerate PeV particles and produce the observed VHE emission, but the synchrotron flux in the X-ray band will be lower than that shown in Figs. \ref{synchrotoron} and \ref{longspectrum}. In this case, the jet should work for longer times to provide the VHE radiation fluxes observed. 

The intermittency of the jet flow of the same average power may also explain the apparently lower X-ray luminosity of V4641 Sgr compared to that in the SS 433 microquasar, which we pointed out at the end of section \ref{Xrays}. The fading time of the turbulent magnetic field in the shock downstream providing the synchrotron radiation is shorter than the lifetime of PeV leptons producing  VHE radiation by the inverse Compton scattering. The intermittent character of the jets in V4641 Sgr is expected from the observed complex X-ray and radio light curves of the source, which is very different from that of SS 433. Deep X-ray observations can therefore bring about interesting information on the evolution of the PeV accelerator.  

\begin{figure}[h]
\includegraphics[scale=0.6]{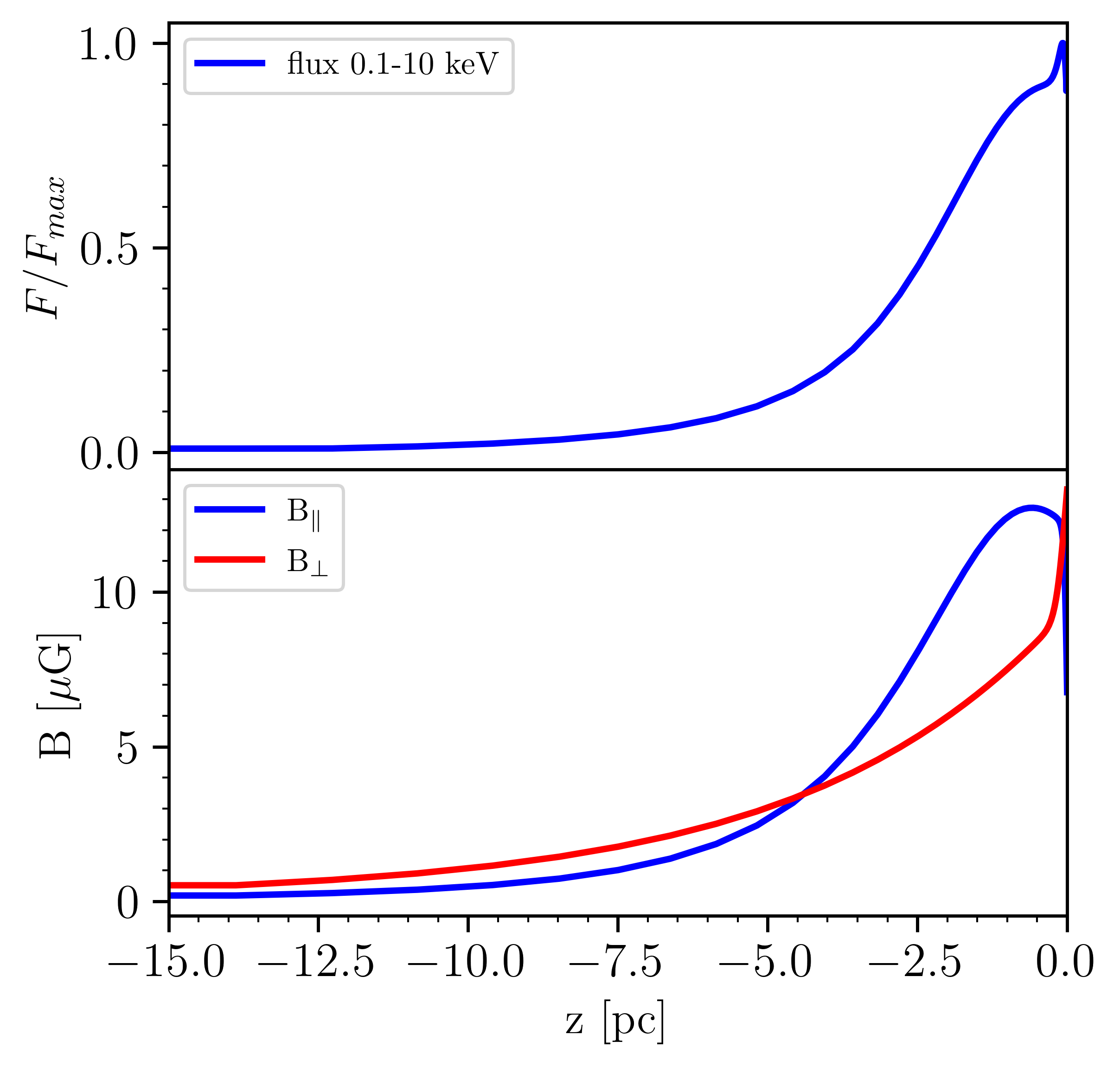}
\caption{Spatial profile of X-ray brightness in relative units for a jet with a turbulent magnetic field of 20 $\mu$G (model jet B). In the lower panel, the profiles of the turbulent magnetic fields parallel (blue) and transverse (red) to the jet direction are shown. These were used to evaluate the predicted spectra. The shock front position is at $z=0$ and the axis is directed to the central source.} 
\label{profile}
\end{figure}

We also modeled the spatial profile of X-ray emission in the vicinity of the shock front for the active accelerator phase. The spatial profile of emissivity of the jet B and the used model of the magnetic field are shown in Figure~\ref{profile}. If the jet shock exists now, then the profile can be tested by deep observations of Chandra or XMM-Newton to constrain the magnetic field structure.

Above, we used the available data from all-sky surveys, which provide the upper limit on the flux in the 2-6 keV band, and, possibly, a detection of hard X-ray flux in the jet B, provided that the sources IGR J18193-2542 and SWIFT BAT J1818.7-2553 are indeed physically associated with the jet.The all-sky survey X-ray data are quite shallow, and SRG/eROSITA has a relatively small effective area in the energy range above 2 keV, meaning that the upper limit presented here is not strongly constraining. However, a factor of 3 increase in sensitivity will already be sufficient to reach the predicted level of the X-ray surface brightness. Deeper dedicated X-ray observations of the extended regions co-spatial with the VHE emission detected by gamma-ray observatories will be able to reveal the activity phase of the accelerator in our model. It is worth noting that the X-ray fluxes from the jets, well below the existing upper limit, might still be compatible with the high detected fluxes of the TeV-PeV radiation.  Within our model, the low X-ray fluxes would indicate that the strong normal recollimation shock disappeared a few hundred years ago - the typical period of decay time of the strong magnetic turbulence in the shock downstream, which is needed to produce the synchrotron X-ray radiation. This would be evidence of intermittent activity of the VHE radiation source. The lifetime of the VHE-energy electrons estimated above is longer than the decay time of the turbulence, so the inverse Compton radiation is emitted for a longer period - a few thousand years.

\subsection{Non-thermal emission from the broad extended wind of the accretion disk}

\begin{figure}[h]
\includegraphics[scale=0.6]{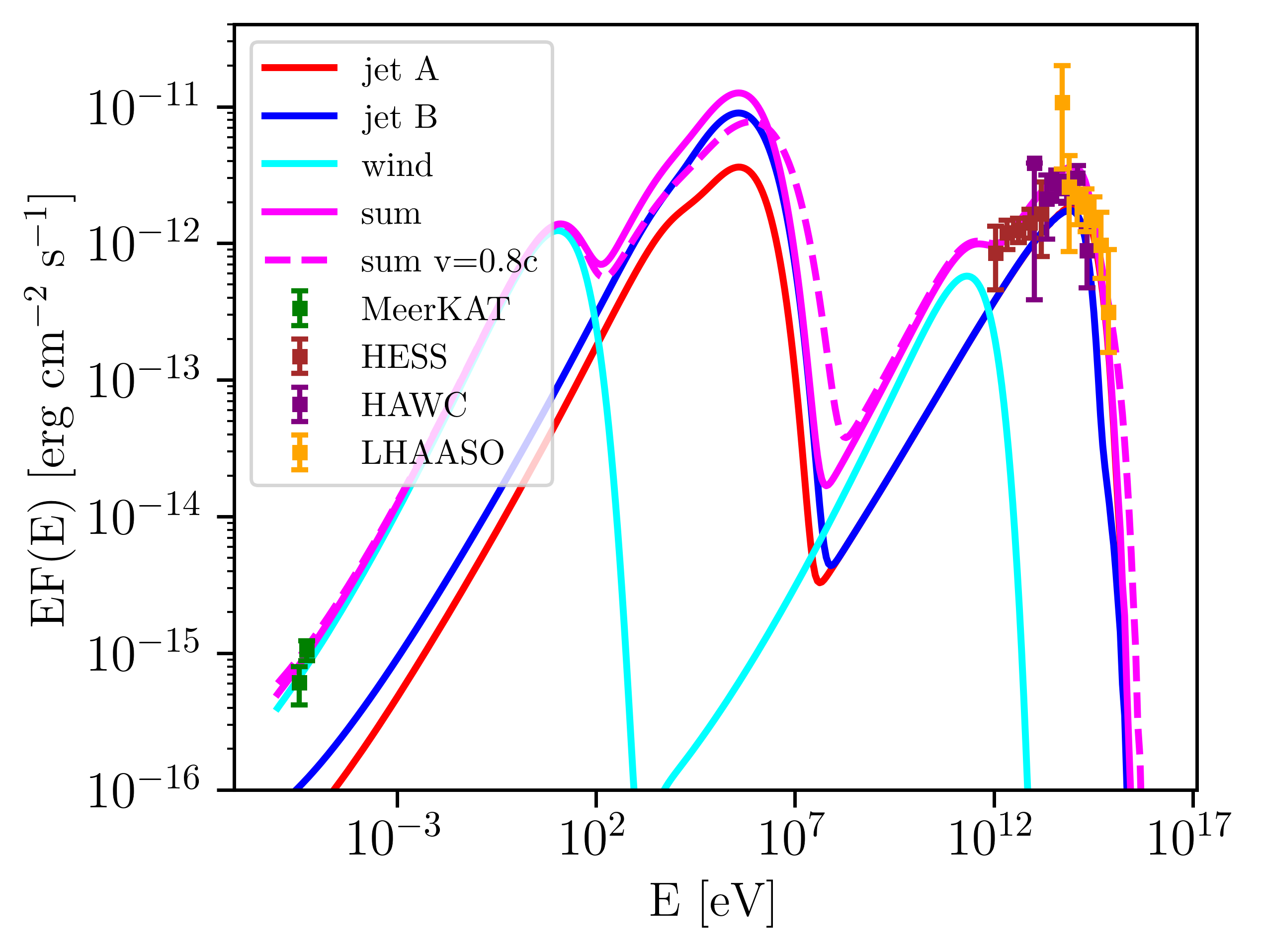}
\caption{Simulated broad band spectrum of electromagnetic radiation from leptons accelerated by the strong shock of the jets of velocity 0.3c and the broad wind of the accretion disk in V4641~Sgr. The available observational data from LHAASO, H.E.S.S., and MeerKAT. The non-thermal radiation simulated for the jets of velocity 0.8c is shown with a dashed line. } 
\label{longspectrum}
\end{figure}

The nearly isotropic non-relativistic wind of velocity 3,000 $\kms$ from the central source creates forward and wind termination shocks (Figs \ref{profile_r_window} and \ref{Scheme}), which accelerate particles. The spectra of accelerated particles were simulated with the non-linear DSA model discussed in section \ref{Kinetic}. The simulated broad band spectrum of the electromagnetic radiation produced by relativistic leptons accelerated in the shocks of the broad wind is shown in Fig. \ref{longspectrum} together with the broad band spectra of both transrelativistic jets with velocities 0.3c. The dashed line in the figure represents the combined spectrum of two jets with velocities 0.8c.

The model reproduces the fluxes of radio emission measured by MeerKAT \citep{MeerKaT2026A&A...706A.283G}. The shape of the large-scale bow-tie structure seen in the L-band MeerKAT wide-field image can be understood in our model. The radio emission 
is produced by isotropic wind terination and forward shocks. the simulated fluxes of the radio emission from the jet-wind interaction regions in Fig. \ref{longspectrum} are dominated by the transrelativistic jets and are well below those of the isotropic wind shocks. The fluxes of synchrotron X-ray radiation from the wind are expected to be low, while the gamma-ray radiation in the GeV-TeV regime could be detected.

\section{Microquasars as galactic cosmic ray sources around the spectral knee}\label{CRs}

The model discussed above presents a scenario that connects the giant power release in the form of the transrelativistic outflows to regions tens of parsecs away from the stellar-mass black hole accreting with the super Eddington rate. 
We show that the observed gamma-ray and non-thermal X-ray emission of both SS~433 and V4641~Sgr within the minimal model with diffusive shock acceleration can be explained as the leptonic radiation of electrons accelerated to PeV energies at the recollimation shocks of transrelativistic jets. Protons and heavier ions are efficiently accelerated in the barionic jets of supercritical kinetic luminosities, as we assumed above, receiving a few percent fraction of the jet power. The accelerated nucleons will leave the accelerator, maintaining the galactic cosmic ray pool and radiating VHE gamma-rays upon traversing enough matter grammage in the source vicinity. The ability of microquasars to accelerate particles to PeV energies makes them excellent candidates for PeV cosmic ray sources among a few other rare classes - e.g. supernova remnants in the special environment in compact stellar clusters \citep[see e.g.][]{BEGO2015MNRAS,2021Univ....7..324C}.  

The existing uncertainties in the current knowledge of the nature of PeV CR sources and their transport in the Galaxy affect the estimations of the PeV proton sources' luminosity needed to explain the observed local CR fluxes at the spectral knee region. By extrapolating the CR escape time from the modeling of the abundance of the secondary nuclei, boron, lithium, and beryllium to the PeV energy range \citet{2019PhRvD..99f3012M} estimated the luminosity of the PeV proton sources to be $2\times 10^{38} \ergs$ for a model with an extended 4 kpc CR halo. On the other hand, local models which attribute the origin of the measured CR fluxes at the spectral knee to the sources located at distances of order 1~kpc and with ages below 1~Myr are discussed as well \citep[see e.g.][]{2026ApJ...999L..13F}. These local models may be energetically less demanding. Our modeling showed that the energetic outflows from both SS~433 and V4641~Sgr microquasars accreting matter at the super-Eddington rate can provide enough power to support even the models with a few kpc CR halos, assuming long enough source activity.    

The estimated global contribution of galactic microquasars to PeV CRs relies on estimations of their jet power transmitted to PeV regime cosmic rays, the number of active microquasars, and their particle acceleration cycles. The expected number of ultra luminous X-ray sources, like SS~433 and V4641~Sgr, in the Galaxy was estimated to be $\sim$ 10 \citep[see e.g.][]{2006MNRAS.370..399B,2007MNRAS.377.1187P}, which is generally consistent with the number of microquasars with the detected VHE gamma-radiation \citep{LHAASO_MQs_2025NSRev..12af496L}. 

The duration of the supercritical accretion phase in these sources is expected to be shorter than 10$^5$ years. PeV particle acceleration in our model is determined by the duty cycle of strong normal shocks in the microquasar jet. Our MHD simulations demonstrated a rather long-lived strong normal shock in the region where the quasi-stationary transrelativistic jet collides with the extended wind termination surface. However, flow inhomogeneities in the jet, which are very likely caused by the time-dependent character of supercritical accretion \citep[see e.g.][]{2026MNRAS.tmp..668F} and are associated with the observed flaring activity of V4641~Sgr, may result in the intermittency of the transrelativistic jet of tens of parsecs in size and appearance/disappearance of the recollimation shocks. 

The amount of time needed to accelerate particles to PeV energies in our nonlinear diffusive shock acceleration model is a few hundred years for a weakly magnetized jet of transrelativistic velocity $\gsim$ 0.3c, and is a few times faster for the jet of 0.8 c velocity. The nonlinear Monte Carlo DSA model included strong MHD turbulence amplified by cosmic ray-driven instabilities. Once the strong normal shock disappears, the magnetic turbulence decays in a few hundred years, and during that period, the VHE energy particles will radiate synchrotron X-rays with intensity fading with time. The lifetime of the PeV leptons against the synchrotron-Compton losses in the jet is also a few hundred years after quenching the shock acceleration process. The MHD simulations presented in Fig. \ref{MHD} showed that the magnitude of the magnetic field in the cocoon is  $\lsim$ 1 $\mu$G. Therefore, the PeV leptons accelerated by the jet shock and escaping to the cocoon will radiate gamma photons via inverse Compton scattering for a few thousand years after the acceleration process stops. In that case, the VHE photon emission region within the cocoon will have a size on the order of 30 pc if the diffusion coefficient of PeV electrons exceeds $10^{28}\diff$. This is consistent with the size of the extended VHE sources detected by HAWC, H.E.S.S., and LHAASO, assuming $\sim$ 6 kpc distance to V4641 Sgr.

\citet{2026arXiv260629830V} recently discussed a possible contribution of the sample of galactic microquasars to the observed fluxes of PeV regime cosmic rays detected at the Earth. Based on the upper limits on the hadronic contribution to the observed gamma-ray flux derived for a sample of galactic microquasars and assuming different efficiencies of the sources, the authors concluded that the observed microquasars contribute less than 37\% of the Galactic cosmic ray spectrum in the PeV regime. 
Our nonlinear DSA model predicted the efficiency of the power transfer from accretion disc outflow to the PeV regime ions to be about a few percent, which is below the maximal efficiency used by \citet{2026arXiv260629830V}. The sources accreting at a supercritical rate demonstrate time variability and flaring activity, which may reduce the PeV particle acceleration periods compared to the full lifetime of the accretor. On the other hand, the variability may affect the estimation of the expected number of galactic microquasars.  A sensitive search for relic VHE gamma-ray sources produced by intermittent activities of the transient galactic microquasars is needed to improve the statistics of PeV sources associated with microquasars.

\section{Summary}\label{Sum}
In this work, we investigated the non-thermal emission from the environment of the galactic microquasar V4641~Sgr powered by supercritical accretion onto a stellar-mass black hole. V4641~Sgr was revealed as a powerful source of PeV-range gamma-ray radiation coming from large (tens of parsecs in size) regions around the black hole. In this paper, the morphology and spectra of V4641~Sgr were simulated within the minimal model 
considering the interaction of the extended broad (accretion disk) wind of velocity $\sim 3000 \kms$ with narrower collimated trans-relativistic polar jets with velocities $\geq$ 0.3c. The total kinetic power in the extended outflows from the accretion disk is $\sim 4 \times 10^{39} \ergs$. It 
was divided between two transrelativistic jets of power $\sim 10^{39} \ergs$ each and the wind power os $\sim 2 \times 10^{39} \ergs$. 
The main conclusions from our analysis are listed below.

(i) MHD modeling of the system revealed the formation of strong recollimation shocks at the positions of the jet collision with the wind termination shock located 12 - 15 pc away from the accreting black hole. The distance is consistent with the observed positions of the bright spots of VHE emission detected in V4641~Sgr, assuming the distance to the source is about 6~kpc.

(ii) The strong shocks accelerate particles by the DSA Fermi mechanism from the jet background plasma and highly amplify the fluctuating magnetic field by cosmic ray-driven instabilities in the jet of initially low magnetization.
According to non-linear kinetic Monte Carlo DSA calculations, both ions and electrons are accelerated to PeV energies, and the jet power transferred to accelerated protons and nuclei of energies above 10 TeV is about 10$^{38}~\ergs$ during the accelerator activity phase, while the power transferred to electrons is at least 100 times less. The magnitude of the turbulent magnetic field amplified by accelerated ions is $\lsim$ 20 $\mu$G in the downstream region of the shock. The spectra of accelerated particles shown in Fig. \ref{pdf_jet} demonstrate a hardening before the cut-off energies, which can be tested or constrained with future sensitive hard X-and gamma-ray observations.     

(iii) The observed VHE radiation up to PeV energies is predicted to be produced by the accelerated leptons via the inverse Compton scattering of CMB photons. The shape of the detected gamma-ray spectrum below 100 TeV is rather hard, which is a potential signature of leptonic radiation within the model. The detected spectra are consistent with the model in which the DSA acceleration phase continues for a few thousand years. The lifetime of accelerated PeV electrons in the cocoon surrounding the jet with magnetic field magnitudes of about 1 $\mu$G is $\gsim$ 10,000 years, which might be much longer than the active phase of the accelerator.       

(iv) Accelerated leptons radiate synchrotron photons up to hard X-ray energies downstream of the shock during the accelerator active phase. In the model, the characteristic sizes of the bright synchrotron knots are set by the magnetic field profile in the shock downstream (see Fig. \ref{profile})  and are about 4-5 pc. The corresponding angular size is about 3\arcmin at the distance of 6 kpc, which is smaller than the reported sizes of VHE emission spots. In turn, the strength of the turbulent magnetic field (Fig.~\ref{profile}) and the level of the synchrotron flux (for a jet of given velocity) depend on the plasma density in the shock upstream. The plasma density in this case is determined by the jet power and its opening angle, which can be derived or constrained from the measured synchrotron flux.

To constrain the model, we analyzed X-ray data from the SRG/eROSITA all-sky survey and archived data from INTEGRAL and Swift BAT observatories. From the SRG/eROSITA short-exposure all-sky survey data, we established an upper limit of $3.7 \times 10^{-12}\,\text{erg}\,\text{s}^{-1}\,\text{cm}^{-2}$ (at the confidence level $3\sigma$) in the energy range of 2 to 6 keV.  The upper limit implies that the X-ray luminosity of the jets in V4641 Sgr in this energy range is below that of in SS 433 while the detected VHE luminosity is higher. 

Hard X-ray fluxes from  IGR J18193-2542 and SWIFT BAT J1818.7-2553 reported by INTEGRAL and Swift BAT (averaged over two decades), might be consistent with the presence of the X-ray jet B of V4641~Sgr, provided the shock in jet B is active now or quit just a hundred years ago.
Deep dedicated observations of the VHE source with X-ray observatories will help to determine whether the accelerator is still active or the current VHE emission is a relic of past activity a few thousand years ago. The intermittent jet activity in V4641 Sgr is expected from the observed X-ray and radio variability of the accreting black hole.    

(v) The prediction of the position of strong normal shocks, which are accelerating PeV energy particles and amplify local magnetic fields, is an essential point of our model. It makes the difference of our scenario with the shear acceleration models, for which the VHE accelerator and emitter positions and the source of the injected particles seem to be less obvious.   

(vi) The extended wind termination shock accelerates multi-TeV regime particles by the DSA mechanism in the segments away from the wind-jet interaction regions. Synchrotron radio emission from the regions downstream of the wind termination shock can explain the morphology and fluxes of the recently reported  L-band by MeerKAT observations \citep{MeerKaT2026A&A...706A.283G}. This directly supports
the assumption of the presence of the broad wind interacting with jets in the minimal model. The simulated fluxes of $\sim$ 100 GeV energy gamma-ray emission from the extended wind of $\sim 4 \times 10^{-13} \enf$ in our Fig. \ref{longspectrum} are below the flux upper limits in the 1 - 300 GeV range established by \citet{Fermi_2025ApJ...984....3Z} from the analysis of 15 yr Fermi-LAT data.

(vii) Hadronic VHE radiation from the jet and cocoon of V4641~Sgr is expected to be below that from the leptons despite the large power transferred to accelerated protons. This is due to the very tenuous plasma within the source. However, the very high energy gamma-rays and neutrinos from the hadronic interactions are expected from the nearby dense clouds, which can be irradiated by accelerated protons (see e.g. \citet{2026A&A...705L...4A,2026JHEAp..5100538C} for a recent discussion).

\section{Acknowledgment}
The authors thank the reviewer for their careful attention to our work and for constructive suggestions on improving the presentation of the results. MHD simulations in this work were performed using the PLUTO code developed by Mignone et al. MHD modeling of the black hole outflows by VIR  with the  Lomonosov-2 computer facilities was supported by RSF grant No. 25-72-20007 (Multiscale nonlinear models of astrophysical sources of high energy radiation). Modeling of very high energy particle radiation by OSM was performed using the resources of the supercomputer center of St. Petersburg Polytechnic University, http://scc.spbstu.ru. IK was supported by the Simons Foundation via the Simons Investigator Award to A. A. Schekochihin.

This work is partly based on observations with the eROSITA telescope onboard \textit{SRG} space observatory. The \textit{SRG} observatory was built by Roskosmos in the interests of the Russian Academy of Sciences represented by its Space Research Institute (IKI) in the framework of the Russian Federal Space Program, with the participation of the Deutsches Zentrum für Luft- und Raumfahrt (DLR). The eROSITA X-ray telescope was built by a consortium of German Institutes led by MPE, and supported by DLR. The \textit{SRG} spacecraft was designed, built, launched, and is operated by the Lavochkin Association and its subcontractors. The science data are downlinked via the Deep Space Network Antennae in Bear Lakes, Ussurijsk, and Baikonur, funded by Roskosmos. 

The development and construction of the eROSITA X-ray instrument was led by MPE, with contributions from the Dr. Karl Remeis Observatory Bamberg $\&$ ECAP (FAU Erlangen-Nuernberg), the University of Hamburg Observatory, the Leibniz Institute for Astrophysics Potsdam (AIP), and the Institute for Astronomy and Astrophysics of the University of Tübingen, with the support of DLR and the Max Planck Society. The Argelander Institute for Astronomy of the University of Bonn and the Ludwig Maximilians Universität Munich also participated in the science preparation for eROSITA. The eROSITA data were processed using the eSASS/NRTA software system developed by the German eROSITA consortium and analyzed using proprietary data reduction software developed by the Russian eROSITA Consortium.


\bibliographystyle{model5-names}

\begin{thebibliography}{73}
\expandafter\ifx\csname natexlab\endcsname\relax\def\natexlab#1{#1}\fi
\providecommand{\url}[1]{\texttt{#1}}
\providecommand{\href}[2]{#2}
\providecommand{\path}[1]{#1}
\providecommand{\DOIprefix}{doi:}
\providecommand{\ArXivprefix}{arXiv:}
\providecommand{\URLprefix}{URL: }
\providecommand{\Pubmedprefix}{pmid:}
\providecommand{\doi}[1]{\href{http://dx.doi.org/#1}{\path{#1}}}
\providecommand{\Pubmed}[1]{\href{pmid:#1}{\path{#1}}}
\providecommand{\bibinfo}[2]{#2}
\ifx\xfnm\relax \def\xfnm[#1]{\unskip,\space#1}\fi
\bibitem[{{Abaroa} et~al.(2026){Abaroa}, {Romero} \& {Bosch-Ramon}}]{2026A&A...705L...4A}
\bibinfo{author}{{Abaroa}, L.}, \bibinfo{author}{{Romero}, G.~E.}, \& \bibinfo{author}{{Bosch-Ramon}, V.} (\bibinfo{year}{2026}).
\newblock \bibinfo{title}{{Microquasar remnants as hidden PeVatrons}}.
\newblock {\it \bibinfo{journal}{\aap}\/},  {\it \bibinfo{volume}{705}\/}, \bibinfo{pages}{L4}. \DOIprefix\doi{10.1051/0004-6361/202557202}. \href{http://arxiv.org/abs/2512.07781}{\tt arXiv:2512.07781}.
\bibitem[{{Abeysekara} et~al.(2018){Abeysekara}, {Albert}, {Alfaro}, {Alvarez}, {{\'A}lvarez}, {Arceo}, {Arteaga-Vel{\'a}zquez}, {Avila Rojas}, {Ayala Solares}, {Belmont-Moreno}, {BenZvi}, {Brisbois}, {Caballero-Mora}, {Capistr{\'a}n}, {Carrami{\~n}ana}, {Casanova}, {Castillo}, {Cotti}, {Cotzomi}, {Couti{\~n}o de Le{\'o}n}, {De Le{\'o}n}, {De la Fuente}, {D{\'\i}az-V{\'e}lez}, {Dichiara}, {Dingus}, {DuVernois}, {Ellsworth}, {Engel}, {Espinoza}, {Fang}, {Fleischhack}, {Fraija}, {Galv{\'a}n-G{\'a}mez}, {Garc{\'\i}a-Gonz{\'a}lez}, {Garfias}, {Gonz{\'a}lez-Mu{\~n}oz}, {Gonz{\'a}lez}, {Goodman}, {Hampel-Arias}, {Harding}, {Hernandez}, {Hinton}, {Hona}, {Hueyotl-Zahuantitla}, {Hui}, {H{\"u}ntemeyer}, {Iriarte}, {Jardin-Blicq}, {Joshi}, {Kaufmann}, {Kar}, {Kunde}, {Lauer}, {Lee}, {Le{\'o}n Vargas}, {Li}, {Linnemann}, {Longinotti}, {Luis-Raya}, {L{\'o}pez-Coto}, {Malone}, {Marinelli}, {Martinez}, {Martinez-Castellanos}, {Mart{\'\i}nez-Castro}, {Matthews}, {Miranda-Romagnoli}, {Moreno}, {Mostaf{\'a}}, {Nayerhoda},
  {Nellen}, {Newbold}, {Nisa}, {Noriega-Papaqui}, {Pretz}, {P{\'e}rez-P{\'e}rez}, {Ren}, {Rho}, {Rivi{\`e}re}, {Rosa-Gonz{\'a}lez}, {Rosenberg}, {Ruiz-Velasco}, {Salesa Greus}, {Sandoval}, {Schneider}, {Schoorlemmer}, {Seglar Arroyo}, {Sinnis}, {Smith}, {Springer}, {Surajbali}, {Taboada}, {Tibolla}, {Tollefson}, {Torres}, {Vianello}, {Villase{\~n}or}, {Weisgarber}, {Werner} et~al.}]{2018Natur.562...82A}
\bibinfo{author}{{Abeysekara}, A.~U.}, \bibinfo{author}{{Albert}, A.}, \bibinfo{author}{{Alfaro}, R.} et~al. (\bibinfo{year}{2018}).
\newblock \bibinfo{title}{{Very-high-energy particle acceleration powered by the jets of the microquasar SS 433}}.
\newblock {\it \bibinfo{journal}{\nat}\/},  {\it \bibinfo{volume}{562}\/}\bibinfo{issue}{(7725)}, \bibinfo{pages}{82--85}. \DOIprefix\doi{10.1038/s41586-018-0565-5}.
\bibitem[{{Acharyya} et~al.(2026){Acharyya}, {Aharonian}, {Ashkar}, {Backes}, {Batzofin}, {Berge}, {Bernl{\"o}hr}, {B{\"o}ttcher}, {Boisson}, {Bolmont}, {Brun}, {Bruno}, {Burger-Scheidlin}, {Bylund}, {Casanova}, {Celic}, {Cerruti}, {Chen}, {Chernyakova}, {Chibueze}, {Chibueze}, {Cornejo}, {Cotter}, {de Assis Scarpin}, {de Bony de Lavergne}, {de Naurois}, {de O{\~n}a Wilhelmi}, {Delgado Giler}, {Devin}, {Djannati-Ata{\"\i}}, {Dmytriiev}, {Egberts}, {Egg}, {Ernenwein}, {Esca{\~n}uela Nieves}, {Fauverge}, {Feijen}, {Filipovic}, {Fontaine}, {Funk}, {Gabici}, {Gallant}, {Glicenstein}, {Glombitza}, {Goswami}, {Grondin}, {Heckmann}, {He{\ss}}, {Hinton}, {Hofmann}, {Holch}, {Holler}, {Jamrozy}, {Jankowsky}, {Jardin-Blicq}, {Jaroschewski}, {Jimeno}, {Jung-Richardt}, {Katarzy{\'n}ski}, {Kerszberg}, {Kh{\'e}lifi}, {Komin}, {Kosack}, {Kostunin}, {Lang}, {Lazarevi{\'c}}, {Lemi{\`e}re}, {Lemoine-Goumard}, {Lenain}, {Liniewicz}, {Luashvili}, {Mackey}, {Malyshev}, {Marandon}, {Mayer}, {Mehta}, {Mitchell}, {Moderski},
  {Mohrmann}, {Montanari}, {Moulin}, {Niemiec}, {Olivera-Nieto}, {Moghadam}, {Panny}, {Parsons}, {Pensec}, {Pichard}, {Preis}, {P{\"u}hlhofer}, {Punch}, {Quirrenbach}, {Reimer}, {Reimer}, {Reis}, {Remy}, {Ren}, {Reville}, {Rieger} et~al.}]{V4641_HESS_2026A&A...706A...8A}
\bibinfo{author}{{Acharyya}, A.}, \bibinfo{author}{{Aharonian}, F.}, \bibinfo{author}{{Ashkar}, H.} et~al. (\bibinfo{year}{2026}).
\newblock \bibinfo{title}{{Constraining the nature of the most extreme Galactic particle accelerator: H.E.S.S. observations of the microquasar V4641 Sgr}}.
\newblock {\it \bibinfo{journal}{\aap}\/},  {\it \bibinfo{volume}{706}\/}, \bibinfo{pages}{A8}. \DOIprefix\doi{10.1051/0004-6361/202557532}. \href{http://arxiv.org/abs/2511.10537}{\tt arXiv:2511.10537}.
\bibitem[{{Aharonian}(2004)}]{2004vhec.book.....A}
\bibinfo{author}{{Aharonian}, F.~A.} (\bibinfo{year}{2004}).
\newblock {\it \bibinfo{title}{{Very high energy cosmic gamma radiation : a crucial window on the extreme Universe}}\/}.
\newblock \DOIprefix\doi{10.1142/4657}.
\bibitem[{{Alfaro} et~al.(2024){Alfaro}, {Alvarez}, {Arteaga-Vel{\'a}zquez}, {Avila Rojas}, {Ayala Solares}, {Babu}, {Belmont-Moreno}, {Caballero-Mora}, {Capistr{\'a}n}, {Carrami{\~n}ana}, {Casanova}, {Cotti}, {Cotzomi}, {Couti{\~n}o de Le{\'o}n}, {De la Fuente}, {Depaoli}, {Di Lalla}, {Diaz Hernandez}, {Dingus}, {DuVernois}, {Durocher}, {D{\'\i}az-V{\'e}lez}, {Engel}, {Espinoza}, {Fan}, {Fang}, {Fraija}, {Fraija}, {Garc{\'\i}a-Gonz{\'a}lez}, {Garfias}, {Gonzalez Mu{\~n}oz}, {Gonz{\'a}lez}, {Goodman}, {Groetsch}, {Harding}, {Herzog}, {Hinton}, {Huang}, {Hueyotl-Zahuantitla}, {H{\"u}ntemeyer}, {Iriarte}, {Joshi}, {Kaufmann}, {Kieda}, {de Le{\'o}n}, {Lee}, {Le{\'o}n Vargas}, {Linnemann}, {Longinotti}, {Luis-Raya}, {Malone}, {Martinez}, {Mart{\'\i}nez-Castro}, {Matthews}, {Miranda-Romagnoli}, {Morales-Soto}, {Moreno}, {Mostaf{\'a}}, {Nayerhoda}, {Nellen}, {Newbold}, {Nisa}, {Noriega-Papaqui}, {Olivera-Nieto}, {Omodei}, {Osorio}, {P{\'e}rez Araujo}, {P{\'e}rez-P{\'e}rez}, {Rho}, {Rosa-Gonz{\'a}lez},
  {Ruiz-Velasco}, {Salazar}, {Salazar-Gallegos}, {Sandoval}, {Schneider}, {Serna-Franco}, {Smith}, {Son}, {Springer}, {Tibolla}, {Tollefson}, {Torres}, {Torres-Escobedo}, {Turner}, {Ure{\~n}a-Mena}, {Varela}, {Villase{\~n}or}, {Wang}, {Watson}, {Willox}, {Yun-C{\'a}rcamo} \& {Zhou}}]{V4641_HAWC_2024Natur.634..557A}
\bibinfo{author}{{Alfaro}, R.}, \bibinfo{author}{{Alvarez}, C.}, \bibinfo{author}{{Arteaga-Vel{\'a}zquez}, J.~C.} et~al. (\bibinfo{year}{2024}).
\newblock \bibinfo{title}{{Ultra-high-energy gamma-ray bubble around microquasar V4641 Sgr}}.
\newblock {\it \bibinfo{journal}{\nat}\/},  {\it \bibinfo{volume}{634}\/}\bibinfo{issue}{(8034)}, \bibinfo{pages}{557--560}. \DOIprefix\doi{10.1038/s41586-024-07995-9}. \href{http://arxiv.org/abs/2410.16117}{\tt arXiv:2410.16117}.
\bibitem[{{Begelman} et~al.(2006){Begelman}, {King} \& {Pringle}}]{2006MNRAS.370..399B}
\bibinfo{author}{{Begelman}, M.~C.}, \bibinfo{author}{{King}, A.~R.}, \& \bibinfo{author}{{Pringle}, J.~E.} (\bibinfo{year}{2006}).
\newblock \bibinfo{title}{{The nature of SS433 and the ultraluminous X-ray sources}}.
\newblock {\it \bibinfo{journal}{\mnras}\/},  {\it \bibinfo{volume}{370}\/}\bibinfo{issue}{(1)}, \bibinfo{pages}{399--404}. \DOIprefix\doi{10.1111/j.1365-2966.2006.10469.x}. \href{http://arxiv.org/abs/astro-ph/0604497}{\tt arXiv:astro-ph/0604497}.
\bibitem[{{Bell}(2004)}]{Bell04}
\bibinfo{author}{{Bell}, A.~R.} (\bibinfo{year}{2004}).
\newblock \bibinfo{title}{{Turbulent amplification of magnetic field and diffusive shock acceleration of cosmic rays}}.
\newblock {\it \bibinfo{journal}{\mnras}\/},  {\it \bibinfo{volume}{353}\/}, \bibinfo{pages}{550--558}. \DOIprefix\doi{10.1111/j.1365-2966.2004.08097.x}.
\bibitem[{{Bird} et~al.(2006){Bird}, {Barlow}, {Bassani}, {Bazzano}, {B{\'e}langer}, {Bodaghee}, {Capitanio}, {Dean}, {Fiocchi}, {Hill}, {Lebrun}, {Malizia}, {Mas-Hesse}, {Molina}, {Moran}, {Renaud}, {Sguera}, {Shaw}, {Stephen}, {Terrier}, {Ubertini}, {Walter}, {Willis} \& {Winkler}}]{2006ApJ...636..765B}
\bibinfo{author}{{Bird}, A.~J.}, \bibinfo{author}{{Barlow}, E.~J.}, \bibinfo{author}{{Bassani}, L.}, \bibinfo{author}{{Bazzano}, A.} et~al. (\bibinfo{year}{2006}).
\newblock \bibinfo{title}{{The Second IBIS/ISGRI Soft Gamma-Ray Survey Catalog}}.
\newblock {\it \bibinfo{journal}{\apj}\/},  {\it \bibinfo{volume}{636}\/}\bibinfo{issue}{(2)}, \bibinfo{pages}{765--776}. \DOIprefix\doi{10.1086/498090}.
\bibitem[{{Blandford} \& {Eichler}(1987)}]{BE87}
\bibinfo{author}{{Blandford}, R.}, \& \bibinfo{author}{{Eichler}, D.} (\bibinfo{year}{1987}).
\newblock \bibinfo{title}{{Particle acceleration at astrophysical shocks: A theory of cosmic ray origin}}.
\newblock {\it \bibinfo{journal}{Phys. Rep.}\/},  {\it \bibinfo{volume}{154}\/}, \bibinfo{pages}{1--75}.
\bibitem[{{Bosch-Ramon} et~al.(2005){Bosch-Ramon}, {Aharonian} \& {Paredes}}]{2005A&A...432..609B}
\bibinfo{author}{{Bosch-Ramon}, V.}, \bibinfo{author}{{Aharonian}, F.~A.}, \& \bibinfo{author}{{Paredes}, J.~M.} (\bibinfo{year}{2005}).
\newblock \bibinfo{title}{{Electromagnetic radiation initiated by hadronic jets from microquasars in the ISM}}.
\newblock {\it \bibinfo{journal}{\aap}\/},  {\it \bibinfo{volume}{432}\/}\bibinfo{issue}{(2)}, \bibinfo{pages}{609--618}. \DOIprefix\doi{10.1051/0004-6361:20041794}. \href{http://arxiv.org/abs/astro-ph/0411508}{\tt arXiv:astro-ph/0411508}.
\bibitem[{{Brinkmann} et~al.(2007){Brinkmann}, {Pratt}, {Rohr}, {Kawai} \& {Burwitz}}]{Brinkmann2007}
\bibinfo{author}{{Brinkmann}, W.}, \bibinfo{author}{{Pratt}, G.~W.}, \bibinfo{author}{{Rohr}, S.}, \bibinfo{author}{{Kawai}, N.}, \& \bibinfo{author}{{Burwitz}, V.} (\bibinfo{year}{2007}).
\newblock \bibinfo{title}{{XMM-Newton observations of the eastern jet of SS 433}}.
\newblock {\it \bibinfo{journal}{\aap}\/},  {\it \bibinfo{volume}{463}\/}\bibinfo{issue}{(2)}, \bibinfo{pages}{611--619}. \DOIprefix\doi{10.1051/0004-6361:20065570}. \href{http://arxiv.org/abs/astro-ph/0610781}{\tt arXiv:astro-ph/0610781}.
\bibitem[{{Bykov} et~al.(2022){Bykov}, {Romansky} \& {Osipov}}]{Bykov2022Univ}
\bibinfo{author}{{Bykov}, A.}, \bibinfo{author}{{Romansky}, V.}, \& \bibinfo{author}{{Osipov}, S.} (\bibinfo{year}{2022}).
\newblock \bibinfo{title}{{Particle Acceleration in Mildly Relativistic Outflows of Fast Energetic Transient Sources}}.
\newblock {\it \bibinfo{journal}{Universe}\/},  {\it \bibinfo{volume}{8}\/}\bibinfo{issue}{(1)}, \bibinfo{pages}{32}. \DOIprefix\doi{10.3390/universe8010032}. \href{http://arxiv.org/abs/2201.11791}{\tt arXiv:2201.11791}.
\bibitem[{{Bykov} et~al.(2015){Bykov}, {Ellison}, {Gladilin} \& {Osipov}}]{BEGO2015MNRAS}
\bibinfo{author}{{Bykov}, A.~M.}, \bibinfo{author}{{Ellison}, D.~C.}, \bibinfo{author}{{Gladilin}, P.~E.}, \& \bibinfo{author}{{Osipov}, S.~M.} (\bibinfo{year}{2015}).
\newblock \bibinfo{title}{{Ultrahard spectra of PeV neutrinos from supernovae in compact star clusters}}.
\newblock {\it \bibinfo{journal}{\mnras}\/},  {\it \bibinfo{volume}{453}\/}, \bibinfo{pages}{113--121}. \DOIprefix\doi{10.1093/mnras/stv1606}. \href{http://arxiv.org/abs/1507.04018}{\tt arXiv:1507.04018}.
\bibitem[{{Bykov} et~al.(2014){Bykov}, {Ellison}, {Osipov} \& {Vladimirov}}]{Bykov3inst2014}
\bibinfo{author}{{Bykov}, A.~M.}, \bibinfo{author}{{Ellison}, D.~C.}, \bibinfo{author}{{Osipov}, S.~M.}, \& \bibinfo{author}{{Vladimirov}, A.~E.} (\bibinfo{year}{2014}).
\newblock \bibinfo{title}{{Magnetic Field Amplification in Nonlinear Diffusive Shock Acceleration Including Resonant and Non-resonant Cosmic-Ray Driven Instabilities}}.
\newblock {\it \bibinfo{journal}{ApJ}\/},  {\it \bibinfo{volume}{789}\/}, \bibinfo{pages}{137}. \DOIprefix\doi{10.1088/0004-637X/789/2/137}. \href{http://arxiv.org/abs/1406.0084}{\tt arXiv:1406.0084}.
\bibitem[{{Bykov} et~al.(2025){Bykov}, {Osipov}, {Romansky}, {Uvarov}, {Churazov} \& {Khabibullin}}]{SS433_min_model_2025PhRvD.112f3017B}
\bibinfo{author}{{Bykov}, A.~M.}, \bibinfo{author}{{Osipov}, S.~M.}, \bibinfo{author}{{Romansky}, V.~I.}, \bibinfo{author}{{Uvarov}, Y.~A.}, \bibinfo{author}{{Churazov}, E.}, \& \bibinfo{author}{{Khabibullin}, I.} (\bibinfo{year}{2025}).
\newblock \bibinfo{title}{{PeV particle acceleration and nonthermal emission in the minimalist model of the extended jets in W50/SS433}}.
\newblock {\it \bibinfo{journal}{\prd}\/},  {\it \bibinfo{volume}{112}\/}\bibinfo{issue}{(6)}, \bibinfo{pages}{063017}. \DOIprefix\doi{10.1103/wws6-3wz8}. \href{http://arxiv.org/abs/2509.09883}{\tt arXiv:2509.09883}.
\bibitem[{{Carpio} et~al.(2026){Carpio}, {Kheirandish} \& {Zhang}}]{2026JHEAp..5100538C}
\bibinfo{author}{{Carpio}, J.~A.}, \bibinfo{author}{{Kheirandish}, A.}, \& \bibinfo{author}{{Zhang}, B.} (\bibinfo{year}{2026}).
\newblock \bibinfo{title}{{Multimessenger emission from very-high-energy black hole-jet systems in the milky way}}.
\newblock {\it \bibinfo{journal}{Journal of High Energy Astrophysics}\/},  {\it \bibinfo{volume}{51}\/}, \bibinfo{pages}{100538}. \DOIprefix\doi{10.1016/j.jheap.2025.100538}. \href{http://arxiv.org/abs/2506.22550}{\tt arXiv:2506.22550}.
\bibitem[{{Churazov} et~al.(2024){Churazov}, {Khabibullin} \& {Bykov}}]{Churazov}
\bibinfo{author}{{Churazov}, E.~M.}, \bibinfo{author}{{Khabibullin}, I.~I.}, \& \bibinfo{author}{{Bykov}, A.~M.} (\bibinfo{year}{2024}).
\newblock \bibinfo{title}{{Minimalist model of the W50/SS433 extended X-ray jet: Anisotropic wind with recollimation shocks}}.
\newblock {\it \bibinfo{journal}{\aap}\/},  {\it \bibinfo{volume}{688}\/}, \bibinfo{pages}{A4}. \DOIprefix\doi{10.1051/0004-6361/202449343}. \href{http://arxiv.org/abs/2401.14770}{\tt arXiv:2401.14770}.
\bibitem[{{Cristofari}(2021)}]{2021Univ....7..324C}
\bibinfo{author}{{Cristofari}, P.} (\bibinfo{year}{2021}).
\newblock \bibinfo{title}{{The Hunt for Pevatrons: The Case of Supernova Remnants}}.
\newblock {\it \bibinfo{journal}{Universe}\/},  {\it \bibinfo{volume}{7}\/}\bibinfo{issue}{(9)}, \bibinfo{pages}{324}. \DOIprefix\doi{10.3390/universe7090324}. \href{http://arxiv.org/abs/2110.07956}{\tt arXiv:2110.07956}.
\bibitem[{{Crumley} et~al.(2019){Crumley}, {Caprioli}, {Markoff} \& {Spitkovsky}}]{2019MNRAS.485.5105C}
\bibinfo{author}{{Crumley}, P.}, \bibinfo{author}{{Caprioli}, D.}, \bibinfo{author}{{Markoff}, S.}, \& \bibinfo{author}{{Spitkovsky}, A.} (\bibinfo{year}{2019}).
\newblock \bibinfo{title}{{Kinetic simulations of mildly relativistic shocks - I. Particle acceleration in high Mach number shocks}}.
\newblock {\it \bibinfo{journal}{\mnras}\/},  {\it \bibinfo{volume}{485}\/}\bibinfo{issue}{(4)}, \bibinfo{pages}{5105--5119}. \DOIprefix\doi{10.1093/mnras/stz232}. \href{http://arxiv.org/abs/1809.10809}{\tt arXiv:1809.10809}.
\bibitem[{{de Angelis} et~al.(2018){de Angelis}, {Tatischeff}, {Grenier}, {McEnery}, {Mallamaci}, {Tavani}, {Oberlack}, {Hanlon}, {Walter}, {Argan}, {von Ballmoos}, {Bulgarelli}, {Bykov}, {Hernanz}, {Kanbach}, {Kuvvetli}, {Pearce}, {Zdziarski}, {Conrad}, {Ghisellini}, {Harding}, {Isern}, {Leising}, {Longo}, {Madejski}, {Martinez}, {Mazziotta}, {Paredes}, {Pohl}, {Rando}, {Razzano}, {Aboudan}, {Ackermann}, {Addazi}, {Ajello}, {Albertus}, {{\'A}lvarez}, {Ambrosi}, {Ant{\'o}n}, {Antonelli}, {Babic}, {Baibussinov}, {Balbo}, {Baldini}, {Balman}, {Bambi}, {Barres de Almeida}, {Barrio}, {Bartels}, {Bastieri}, {Bednarek}, {Bernard}, {Bernardini}, {Bernasconi}, {Bertucci}, {Biland}, {Bissaldi}, {Boettcher}, {Bonvicini}, {Bosch-Ramon}, {Bottacini}, {Bozhilov}, {Bretz}, {Branchesi}, {Brdar}, {Bringmann}, {Brogna}, {Budtz J{\o}rgensen}, {Busetto}, {Buson}, {Busso}, {Caccianiga}, {Camera}, {Campana}, {Caraveo}, {Cardillo}, {Carlson}, {Celestin}, {Cerme{\~n}o}, {Chen}, {Cheung}, {Churazov}, {Ciprini}, {Coc},
  {Colafrancesco}, {Coleiro}, {Collmar}, {Coppi}, {Curado da Silva}, {Cutini}, {D'Ammando}, {de Lotto}, {de Martino}, {De Rosa}, {Del Santo}, {Delgado}, {Diehl}, {Dietrich}, {Dolgov} et~al.}]{2018JHEAp..19....1D}
\bibinfo{author}{{de Angelis}, A.}, \bibinfo{author}{{Tatischeff}, V.}, \bibinfo{author}{{Grenier}, I.~A.}  et~al. (\bibinfo{year}{2018}).
\newblock \bibinfo{title}{{Science with e-ASTROGAM. A space mission for MeV-GeV gamma-ray astrophysics}}.
\newblock {\it \bibinfo{journal}{Journal of High Energy Astrophysics}\/},  {\it \bibinfo{volume}{19}\/}, \bibinfo{pages}{1--106}. \DOIprefix\doi{10.1016/j.jheap.2018.07.001}. \href{http://arxiv.org/abs/1711.01265}{\tt arXiv:1711.01265}.
\bibitem[{{Dmytriiev} et~al.(2025){Dmytriiev}, {van der Merwe} \& {B{\"o}ttcher}}]{2025arXiv251202574D}
\bibinfo{author}{{Dmytriiev}, A.}, \bibinfo{author}{{van der Merwe}, F.}, \& \bibinfo{author}{{B{\"o}ttcher}, M.} (\bibinfo{year}{2025}).
\newblock \bibinfo{title}{{Modeling Particle Acceleration and MWL Emission of a PeVatron Microquasar V4641 Sgr}}.
\newblock {\it \bibinfo{journal}{arXiv e-prints}\/},  (p. \bibinfo{pages}{arXiv:2512.02574}). \DOIprefix\doi{10.48550/arXiv.2512.02574}. \href{http://arxiv.org/abs/2512.02574}{\tt arXiv:2512.02574}.
\bibitem[{{Dmytriiev} et~al.(2026){Dmytriiev}, {van der Merwe} \& {B{\"o}ttcher}}]{2026arXiv260309394D}
\bibinfo{author}{{Dmytriiev}, A.}, \bibinfo{author}{{van der Merwe}, F.}, \& \bibinfo{author}{{B{\"o}ttcher}, M.} (\bibinfo{year}{2026}).
\newblock \bibinfo{title}{{Beyond Fermi-II: Intermittent Particle Acceleration by Relativistic Turbulence in Astrophysical Plasmas}}.
\newblock {\it \bibinfo{journal}{arXiv e-prints}\/},  (p. \bibinfo{pages}{arXiv:2603.09394}). \DOIprefix\doi{10.48550/arXiv.2603.09394}. \href{http://arxiv.org/abs/2603.09394}{\tt arXiv:2603.09394}.
\bibitem[{{Dubus} et~al.(2010){Dubus}, {Cerutti} \& {Henri}}]{2010MNRAS.404L..55D}
\bibinfo{author}{{Dubus}, G.}, \bibinfo{author}{{Cerutti}, B.}, \& \bibinfo{author}{{Henri}, G.} (\bibinfo{year}{2010}).
\newblock \bibinfo{title}{{The relativistic jet of Cygnus X-3 in gamma-rays}}.
\newblock {\it \bibinfo{journal}{\mnras}\/},  {\it \bibinfo{volume}{404}\/}\bibinfo{issue}{(1)}, \bibinfo{pages}{L55--L59}. \DOIprefix\doi{10.1111/j.1745-3933.2010.00834.x}. \href{http://arxiv.org/abs/1002.3888}{\tt arXiv:1002.3888}.
\bibitem[{{Fang} \& {Halzen}(2026)}]{2026ApJ...999L..13F}
\bibinfo{author}{{Fang}, K.}, \& \bibinfo{author}{{Halzen}, F.} (\bibinfo{year}{2026}).
\newblock \bibinfo{title}{{The Cosmic-Ray Knee as a Local Signature of Nearby PeVatrons}}.
\newblock {\it \bibinfo{journal}{\apjl}\/},  {\it \bibinfo{volume}{999}\/}\bibinfo{issue}{(1)}, \bibinfo{pages}{L13}. \DOIprefix\doi{10.3847/2041-8213/ae433a}. \href{http://arxiv.org/abs/2601.05435}{\tt arXiv:2601.05435}.
\bibitem[{{Fragile} et~al.(2026){Fragile}, {Middleton}, {Brasseur}, {Bollimpalli} \& {Smith}}]{2026MNRAS.tmp..668F}
\bibinfo{author}{{Fragile}, P.~C.}, \bibinfo{author}{{Middleton}, M.~J.}, \bibinfo{author}{{Brasseur}, B.}, \bibinfo{author}{{Bollimpalli}, D.~A.}, \& \bibinfo{author}{{Smith}, Z.} (\bibinfo{year}{2026}).
\newblock \bibinfo{title}{{The nature of tilted supercritical accretion discs}}.
\newblock {\it \bibinfo{journal}{\mnras}\/}, . \DOIprefix\doi{10.1093/mnras/stag711}. \href{http://arxiv.org/abs/2604.11794}{\tt arXiv:2604.11794}.
\bibitem[{{Ginzburg} \& {Syrovatskii}(1965)}]{1965ARA&A...3..297G}
\bibinfo{author}{{Ginzburg}, V.~L.}, \& \bibinfo{author}{{Syrovatskii}, S.~I.} (\bibinfo{year}{1965}).
\newblock \bibinfo{title}{{Cosmic Magnetobremsstrahlung (synchrotron Radiation)}}.
\newblock {\it \bibinfo{journal}{\araa}\/},  {\it \bibinfo{volume}{3}\/}, \bibinfo{pages}{297}. \DOIprefix\doi{10.1146/annurev.aa.03.090165.001501}.
\bibitem[{{Grollimund} et~al.(2026){Grollimund}, {Corbel}, {Fender}, {Matthews}, {Heywood}, {Cowie}, {Hughes}, {Carotenuto}, {Motta} \& {Woudt}}]{MeerKaT2026A&A...706A.283G}
\bibinfo{author}{{Grollimund}, N.}, \bibinfo{author}{{Corbel}, S.}, \bibinfo{author}{{Fender}, R.}, \bibinfo{author}{{Matthews}, J.~H.}, \bibinfo{author}{{Heywood}, I.}, \bibinfo{author}{{Cowie}, F.~J.}, \bibinfo{author}{{Hughes}, A.~K.}, \bibinfo{author}{{Carotenuto}, F.}, \bibinfo{author}{{Motta}, S.~E.}, \& \bibinfo{author}{{Woudt}, P.} (\bibinfo{year}{2026}).
\newblock \bibinfo{title}{{Large-scale radio bubbles around the black hole transient V4641 Sgr}}.
\newblock {\it \bibinfo{journal}{\aap}\/},  {\it \bibinfo{volume}{706}\/}, \bibinfo{pages}{A283}. \DOIprefix\doi{10.1051/0004-6361/202557124}. \href{http://arxiv.org/abs/2601.03140}{\tt arXiv:2601.03140}.
\bibitem[{{H.~E.~S.~S. Collaboration} et~al.(2024){H.~E.~S.~S. Collaboration}, {Aharonian}, {Ait Benkhali}, {Aschersleben}, {Ashkar}, {Backes}, {Barbosa Martins}, {Batzofin}, {Becherini}, {Berge}, {Bernl{\"o}hr}, {Bi}, {B{\"o}ttcher}, {Boisson}, {Bolmont}, {de Lavergne}, {Borowska}, {Bouyahiaoui}, {Breuhaus}, {Brose}, {Brown}, {Brun}, {Bruno}, {Bulik}, {Burger-Scheidlin}, {Caroff}, {Casanova}, {Cecil}, {Celic}, {Cerruti}, {Chand}, {Chandra}, {Chen}, {Chibueze}, {Chibueze}, {Cotter}, {Dai}, {Mbarubucyeye}, {Djannati-Ata{\"\i}}, {Dmytriiev}, {Doroshenko}, {Egberts}, {Einecke}, {Ernenwein}, {Filipovic}, {Fontaine}, {F{\"u}{\ss}ling}, {Funk}, {Gabici}, {Ghafourizadeh}, {Giavitto}, {Glawion}, {Glicenstein}, {Grolleron}, {Haerer}, {Hinton}, {Hofmann}, {Holch}, {Holler}, {Horns}, {Jamrozy}, {Jankowsky}, {Jardin-Blicq}, {Joshi}, {Jung-Richardt}, {Kasai}, {Katarzy{\'n}ski}, {Khatoon}, {Kh{\'e}lifi}, {Klepser}, {Klu{\'z}niak}, {Komin}, {Kosack}, {Kostunin}, {Kundu}, {Lang}, {Le Stum}, {Leitl}, {Lemi{\`e}re}, {Lenain},
  {Leuschner}, {Lohse}, {Luashvili}, {Lypova}, {Mackey}, {Malyshev}, {Malyshev}, {Marandon}, {Marchegiani}, {Marcowith}, {Mart{\'\i}-Devesa}, {Marx}, {Mehta}, {Mitchell}, {Moderski}, {Mohrmann}, {Montanari}, {Moulin}, {Murach} et~al.}]{HESS2024SS433}
\bibinfo{author}{{H.~E.~S.~S. Collaboration}}, \bibinfo{author}{{Aharonian}, F.}, \bibinfo{author}{{Ait Benkhali}, F.}, \bibinfo{author}{{Aschersleben}, J.} et~al. (\bibinfo{year}{2024}).
\newblock \bibinfo{title}{{Acceleration and transport of relativistic electrons in the jets of the microquasar SS 433}}.
\newblock {\it \bibinfo{journal}{Science}\/},  {\it \bibinfo{volume}{383}\/}\bibinfo{issue}{(6681)}, \bibinfo{pages}{402--406}. \DOIprefix\doi{10.1126/science.adi2048}. \href{http://arxiv.org/abs/2401.16019}{\tt arXiv:2401.16019}.
\bibitem[{{in't Zand} et~al.(2000){in't Zand}, {Kuulkers}, {Bazzano}, {Cornelisse}, {Cocchi}, {Heise}, {Muller}, {Natalucci}, {Smith} \& {Ubertini}}]{2000A&A...357..520I}
\bibinfo{author}{{in't Zand}, J.~J.~M.}, \bibinfo{author}{{Kuulkers}, E.}, \bibinfo{author}{{Bazzano}, A.} et~al.(\bibinfo{year}{2000}).
\newblock \bibinfo{title}{{BeppoSAX observations of the nearby low-mass X-ray binary and fast transient SAX J1819.3-2525}}.
\newblock {\it \bibinfo{journal}{\aap}\/},  {\it \bibinfo{volume}{357}\/}, \bibinfo{pages}{520--526}. \DOIprefix\doi{10.48550/arXiv.astro-ph/0001110}. \href{http://arxiv.org/abs/astro-ph/0001110}{\tt arXiv:astro-ph/0001110}.
\bibitem[{{Jikei} et~al.(2026){Jikei}, {Groselj} \& {Sironi}}]{transrel_sironi_2026ApJ...998..149J}
\bibinfo{author}{{Jikei}, T.}, \bibinfo{author}{{Groselj}, D.}, \& \bibinfo{author}{{Sironi}, L.} (\bibinfo{year}{2026}).
\newblock \bibinfo{title}{{Magnetic Field Amplification and Particle Acceleration in Weakly Magnetized Transrelativistic Electron-Ion Shocks}}.
\newblock {\it \bibinfo{journal}{\apj}\/},  {\it \bibinfo{volume}{998}\/}\bibinfo{issue}{(1)}, \bibinfo{pages}{149}. \DOIprefix\doi{10.3847/1538-4357/ae3723}. \href{http://arxiv.org/abs/2512.03169}{\tt arXiv:2512.03169}.
\bibitem[{{Jones}(1968)}]{Jones68}
\bibinfo{author}{{Jones}, F.~C.} (\bibinfo{year}{1968}).
\newblock \bibinfo{title}{{Calculated Spectrum of Inverse-Compton-Scattered Photons}}.
\newblock {\it \bibinfo{journal}{Physical Review}\/},  {\it \bibinfo{volume}{167}\/}, \bibinfo{pages}{1159--1169}.
\bibitem[{{Jones} \& {Ellison}(1991)}]{JE91}
\bibinfo{author}{{Jones}, F.~C.}, \& \bibinfo{author}{{Ellison}, D.~C.} (\bibinfo{year}{1991}).
\newblock \bibinfo{title}{{The plasma physics of shock acceleration}}.
\newblock {\it \bibinfo{journal}{Space Science Reviews}\/},  {\it \bibinfo{volume}{58}\/}, \bibinfo{pages}{259--346}.
\bibitem[{{Kaaret} et~al.(2024){Kaaret}, {Ferrazzoli}, {Silvestri}, {Negro}, {Manfreda}, {Wu}, {Costa}, {Soffitta}, {Safi-Harb}, {Poutanen}, {Veledina}, {Di Marco}, {Slane}, {Bianchi}, {Ingram}, {Romani}, {Cibrario}, {Mac Intyre}, {Mikusincov{\'a}}, {Ratheesh}, {Steiner}, {Svoboda}, {Tugliani}, {Agudo}, {Antonelli}, {Bachetti}, {Baldini}, {Baumgartner}, {Bellazzini}, {Bongiorno}, {Bonino}, {Brez}, {Bucciantini}, {Capitanio}, {Castellano}, {Cavazzuti}, {Chen}, {Ciprini}, {De Rosa}, {Del Monte}, {Di Gesu}, {Di Lalla}, {Donnarumma}, {Doroshenko}, {Dov{\v{c}}iak}, {Ehlert}, {Enoto}, {Evangelista}, {Fabiani}, {Garc{\'\i}a}, {Gunji}, {Hayashida}, {Heyl}, {Iwakiri}, {Jorstad}, {Karas}, {Kislat}, {Kitaguchi}, {Kolodziejczak}, {Krawczynski}, {La Monaca}, {Latronico}, {Liodakis}, {Maldera}, {Marin}, {Marinucci}, {Marscher}, {Marshall}, {Massaro}, {Matt}, {Mitsuishi}, {Mizuno}, {Muleri}, {Ng}, {O'Dell}, {Omodei}, {Oppedisano}, {Papitto}, {Pavlov}, {Peirson}, {Perri}, {Pesce-Rollins}, {Petrucci}, {Pilia}, {Possenti},
  {Puccetti}, {Ramsey}, {Rankin}, {Roberts}, {Sgr{\`o}}, {Spandre}, {Swartz}, {Tamagawa}, {Tavecchio}, {Taverna}, {Tawara}, {Tennant}, {Thomas}, {Tombesi} et~al.}]{IXPE2024SS433}
\bibinfo{author}{{Kaaret}, P.}, \bibinfo{author}{{Ferrazzoli}, R.}, \bibinfo{author}{{Silvestri}, S.}  et~al. (\bibinfo{year}{2024}).
\newblock \bibinfo{title}{{X-Ray Polarization of the Eastern Lobe of SS 433}}.
\newblock {\it \bibinfo{journal}{\apjl}\/},  {\it \bibinfo{volume}{961}\/}\bibinfo{issue}{(1)}, \bibinfo{pages}{L12}. \DOIprefix\doi{10.3847/2041-8213/ad103b}. \href{http://arxiv.org/abs/2311.16313}{\tt arXiv:2311.16313}.
\bibitem[{{Kaiser} et~al.(2000){Kaiser}, {Sunyaev} \& {Spruit}}]{2000A&A...356..975K}
\bibinfo{author}{{Kaiser}, C.~R.}, \bibinfo{author}{{Sunyaev}, R.}, \& \bibinfo{author}{{Spruit}, H.~C.} (\bibinfo{year}{2000}).
\newblock \bibinfo{title}{{Internal shock model for microquasars}}.
\newblock {\it \bibinfo{journal}{\aap}\/},  {\it \bibinfo{volume}{356}\/}, \bibinfo{pages}{975--988}. \DOIprefix\doi{10.48550/arXiv.astro-ph/0001501}. \href{http://arxiv.org/abs/astro-ph/0001501}{\tt arXiv:astro-ph/0001501}.
\bibitem[{{Khangulyan} et~al.(2024){Khangulyan}, {Bosch-Ramon} \& {Hadasch}}]{2024JHEAp..43...93K}
\bibinfo{author}{{Khangulyan}, D.}, \bibinfo{author}{{Bosch-Ramon}, V.}, \& \bibinfo{author}{{Hadasch}, D.} (\bibinfo{year}{2024}).
\newblock \bibinfo{title}{{Non-thermal emission from microquasar jets: The case of GRS 1915+105}}.
\newblock {\it \bibinfo{journal}{Journal of High Energy Astrophysics}\/},  {\it \bibinfo{volume}{43}\/}, \bibinfo{pages}{93--104}. \DOIprefix\doi{10.1016/j.jheap.2024.06.006}.
\bibitem[{{Kimura} et~al.(2019){Kimura}, {Tomida} \& {Murase}}]{2019MNRAS.485..163K}
\bibinfo{author}{{Kimura}, S.~S.}, \bibinfo{author}{{Tomida}, K.}, \& \bibinfo{author}{{Murase}, K.} (\bibinfo{year}{2019}).
\newblock \bibinfo{title}{{Acceleration and escape processes of high-energy particles in turbulence inside hot accretion flows}}.
\newblock {\it \bibinfo{journal}{\mnras}\/},  {\it \bibinfo{volume}{485}\/}\bibinfo{issue}{(1)}, \bibinfo{pages}{163--178}. \DOIprefix\doi{10.1093/mnras/stz329}. \href{http://arxiv.org/abs/1812.03901}{\tt arXiv:1812.03901}.
\bibitem[{{Kleimenov} et~al.(2025){Kleimenov}, {Neronov}, {Oikonomou} \& {Semikoz}}]{V4641_lepton_2025arXiv251213578K}
\bibinfo{author}{{Kleimenov}, M.}, \bibinfo{author}{{Neronov}, A.}, \bibinfo{author}{{Oikonomou}, F.}, \& \bibinfo{author}{{Semikoz}, D.} (\bibinfo{year}{2025}).
\newblock \bibinfo{title}{{Leptonic and Hadronic Models of High-energy Nebula Around V4641 Sgr}}.
\newblock {\it \bibinfo{journal}{arXiv e-prints}\/},  (p. \bibinfo{pages}{arXiv:2512.13578}). \DOIprefix\doi{10.48550/arXiv.2512.13578}. \href{http://arxiv.org/abs/2512.13578}{\tt arXiv:2512.13578}.
\bibitem[{{Koljonen} \& {Tomsick}(2020)}]{2020A&A...639A..13K}
\bibinfo{author}{{Koljonen}, K.~I.~I.}, \& \bibinfo{author}{{Tomsick}, J.~A.} (\bibinfo{year}{2020}).
\newblock \bibinfo{title}{{The obscured X-ray binaries V404 Cyg, Cyg X-3, V4641 Sgr, and GRS 1915+105}}.
\newblock {\it \bibinfo{journal}{\aap}\/},  {\it \bibinfo{volume}{639}\/}, \bibinfo{pages}{A13}. \DOIprefix\doi{10.1051/0004-6361/202037882}. \href{http://arxiv.org/abs/2004.08536}{\tt arXiv:2004.08536}.
\bibitem[{{Komissarov} \& {Falle}(1997)}]{KomissarovFalle1997}
\bibinfo{author}{{Komissarov}, S.~S.}, \& \bibinfo{author}{{Falle}, S.~A.~E.~G.} (\bibinfo{year}{1997}).
\newblock \bibinfo{title}{{Simulations of Superluminal Radio Sources}}.
\newblock {\it \bibinfo{journal}{\mnras}\/},  {\it \bibinfo{volume}{288}\/}\bibinfo{issue}{(4)}, \bibinfo{pages}{833--848}. \DOIprefix\doi{10.1093/mnras/288.4.833}.
\bibitem[{{Krivonos} et~al.(2022){Krivonos}, {Sazonov}, {Kuznetsova}, {Lutovinov}, {Mereminskiy} \& {Tsygankov}}]{2022MNRAS.510.4796K}
\bibinfo{author}{{Krivonos}, R.~A.}, \bibinfo{author}{{Sazonov}, S.~Y.}, \bibinfo{author}{{Kuznetsova}, E.~A.}, \bibinfo{author}{{Lutovinov}, A.~A.}, \bibinfo{author}{{Mereminskiy}, I.~A.}, \& \bibinfo{author}{{Tsygankov}, S.~S.} (\bibinfo{year}{2022}).
\newblock \bibinfo{title}{{INTEGRAL/IBIS 17-yr hard X-ray all-sky survey}}.
\newblock {\it \bibinfo{journal}{\mnras}\/},  {\it \bibinfo{volume}{510}\/}\bibinfo{issue}{(4)}, \bibinfo{pages}{4796--4807}. \DOIprefix\doi{10.1093/mnras/stab3751}. \href{http://arxiv.org/abs/2111.02996}{\tt arXiv:2111.02996}.
\bibitem[{{Lebrun} et~al.(2003){Lebrun}, {Leray}, {Lavocat}, {Cr{\'e}tolle}, {Arqu{\`e}s}, {Blondel}, {Bonnin}, {Bou{\`e}re}, {Cara}, {Chaleil}, {Daly}, {Desages}, {Dzitko}, {Horeau}, {Laurent}, {Limousin}, {Mathy}, {Mauguen}, {Meignier}, {Molini{\'e}}, {Poindron}, {Rouger}, {Sauvageon} \& {Tourrette}}]{2003A&A...411L.141L}
\bibinfo{author}{{Lebrun}, F.}, \bibinfo{author}{{Leray}, J.~P.}, \bibinfo{author}{{Lavocat}, P.} et~al.(\bibinfo{year}{2003}).
\newblock \bibinfo{title}{{ISGRI: The INTEGRAL Soft Gamma-Ray Imager}}.
\newblock {\it \bibinfo{journal}{\aap}\/},  {\it \bibinfo{volume}{411}\/}, \bibinfo{pages}{L141--L148}. \DOIprefix\doi{10.1051/0004-6361:20031367}. \href{http://arxiv.org/abs/astro-ph/0310362}{\tt arXiv:astro-ph/0310362}.
\bibitem[{{Lemoine} \& {Rieger}(2025)}]{2025A&A...697A.124L}
\bibinfo{author}{{Lemoine}, M.}, \& \bibinfo{author}{{Rieger}, F.} (\bibinfo{year}{2025}).
\newblock \bibinfo{title}{{Neutrinos from stochastic acceleration in black hole environments}}.
\newblock {\it \bibinfo{journal}{\aap}\/},  {\it \bibinfo{volume}{697}\/}, \bibinfo{pages}{A124}. \DOIprefix\doi{10.1051/0004-6361/202453296}. \href{http://arxiv.org/abs/2412.01457}{\tt arXiv:2412.01457}.
\bibitem[{{Liu} et~al.(2026){Liu}, {Shao} \& {Nie}}]{V4641_bin_accr26}
\bibinfo{author}{{Liu}, R.-Y.}, \bibinfo{author}{{Shao}, Y.}, \& \bibinfo{author}{{Nie}, Y.-D.} (\bibinfo{year}{2026}).
\newblock \bibinfo{title}{{The long-term accretion luminosity of V4641 Sgr through binary evolution simulations: implications for its ultrahigh-energy gamma-ray emission}}.
\newblock {\it \bibinfo{journal}{arXiv e-prints}\/},  (p. \bibinfo{pages}{arXiv:2603.16714}). \DOIprefix\doi{10.48550/arXiv.2603.16714}. \href{http://arxiv.org/abs/2603.16714}{\tt arXiv:2603.16714}.
\bibitem[{{Mart{\'\i}} \& {Luque-Escamilla}(2026)}]{Marti_V4641_VLA_jets26}
\bibinfo{author}{{Mart{\'\i}}, J.}, \& \bibinfo{author}{{Luque-Escamilla}, P.~L.} (\bibinfo{year}{2026}).
\newblock \bibinfo{title}{{Alignment of radio jets in the microquasar V4641 Sagittarii with its high-energy structures}}.
\newblock {\it \bibinfo{journal}{\mnras}\/},  {\it \bibinfo{volume}{545}\/}\bibinfo{issue}{(3)}, \bibinfo{pages}{staf2104}. \DOIprefix\doi{10.1093/mnras/staf2104}. \href{http://arxiv.org/abs/2511.19695}{\tt arXiv:2511.19695}.
\bibitem[{{Mart{\'\i}} et~al.(1997){Mart{\'\i}}, {M{\"u}ller}, {Font}, {Ib{\'a}{\~n}ez} \& {Marquina}}]{recoll_1997ApJ...479..151M}
\bibinfo{author}{{Mart{\'\i}}, J.~M.}, \bibinfo{author}{{M{\"u}ller}, E.}, \bibinfo{author}{{Font}, J.~A.}, \bibinfo{author}{{Ib{\'a}{\~n}ez}, J.~M.~Z.}, \& \bibinfo{author}{{Marquina}, A.} (\bibinfo{year}{1997}).
\newblock \bibinfo{title}{{Morphology and Dynamics of Relativistic Jets}}.
\newblock {\it \bibinfo{journal}{\apj}\/},  {\it \bibinfo{volume}{479}\/}\bibinfo{issue}{(1)}, \bibinfo{pages}{151--163}. \DOIprefix\doi{10.1086/303842}.
\bibitem[{{Middleton} et~al.(2021){Middleton}, {Walton}, {Alston}, {Dauser}, {Eikenberry}, {Jiang}, {Fabian}, {Fuerst}, {Brightman}, {Marshall}, {Parker}, {Pinto}, {Harrison}, {Bachetti}, {Altamirano}, {Bird}, {Perez}, {Miller-Jones}, {Charles}, {Boggs}, {Christensen}, {Craig}, {Forster}, {Grefenstette}, {Hailey}, {Madsen}, {Stern} \& {Zhang}}]{2021MNRAS.506.1045M}
\bibinfo{author}{{Middleton}, M.~J.}, \bibinfo{author}{{Walton}, D.~J.}, \bibinfo{author}{{Alston}, W.} et~al. (\bibinfo{year}{2021}).
\newblock \bibinfo{title}{{NuSTAR reveals the hidden nature of SS433}}.
\newblock {\it \bibinfo{journal}{\mnras}\/},  {\it \bibinfo{volume}{506}\/}\bibinfo{issue}{(1)}, \bibinfo{pages}{1045--1058}. \DOIprefix\doi{10.1093/mnras/stab1280}. \href{http://arxiv.org/abs/1810.10518}{\tt arXiv:1810.10518}.
\bibitem[{{Mignone} et~al.(2007){Mignone}, {Bodo}, {Massaglia}, {Matsakos}, {Tesileanu}, {Zanni} \& {Ferrari}}]{Mignone2007}
\bibinfo{author}{{Mignone}, A.}, \bibinfo{author}{{Bodo}, G.}, \bibinfo{author}{{Massaglia}, S.} et~al. (\bibinfo{year}{2007}).
\newblock \bibinfo{title}{{PLUTO: A Numerical Code for Computational Astrophysics}}.
\newblock {\it \bibinfo{journal}{\apjs}\/},  {\it \bibinfo{volume}{170}\/}\bibinfo{issue}{(1)}, \bibinfo{pages}{228--242}. \DOIprefix\doi{10.1086/513316}. \href{http://arxiv.org/abs/astro-ph/0701854}{\tt arXiv:astro-ph/0701854}.
\bibitem[{{Mu{\~n}oz-Darias} et~al.(2026){Mu{\~n}oz-Darias}, {D{\'\i}az Trigo}, {Done}, {Ponti} \& {Tomaru}}]{accr_disc_winds_SSRv26}
\bibinfo{author}{{Mu{\~n}oz-Darias}, T.}, \bibinfo{author}{{D{\'\i}az Trigo}, M.}, \bibinfo{author}{{Done}, C.}, \bibinfo{author}{{Ponti}, G.}, \& \bibinfo{author}{{Tomaru}, R.} (\bibinfo{year}{2026}).
\newblock \bibinfo{title}{{Accretion Disc Winds in X-ray Binaries}}.
\newblock {\it \bibinfo{journal}{\ssr}\/},  {\it \bibinfo{volume}{222}\/}\bibinfo{issue}{(4)}, \bibinfo{pages}{39}. \DOIprefix\doi{10.1007/s11214-026-01292-9}. \href{http://arxiv.org/abs/2601.05319}{\tt arXiv:2601.05319}.
\bibitem[{{Mu{\~n}oz-Darias} et~al.(2018){Mu{\~n}oz-Darias}, {Torres} \& {Garcia}}]{V4641wind18}
\bibinfo{author}{{Mu{\~n}oz-Darias}, T.}, \bibinfo{author}{{Torres}, M. A.~P.}, \& \bibinfo{author}{{Garcia}, M.~R.} (\bibinfo{year}{2018}).
\newblock \bibinfo{title}{{The low-luminosity accretion disc wind of the black hole transient V4641 Sagittarii}}.
\newblock {\it \bibinfo{journal}{\mnras}\/},  {\it \bibinfo{volume}{479}\/}\bibinfo{issue}{(3)}, \bibinfo{pages}{3987--3995}. \DOIprefix\doi{10.1093/mnras/sty1711}. \href{http://arxiv.org/abs/1806.10881}{\tt arXiv:1806.10881}.
\bibitem[{{Murase} \& {Fukugita}(2019)}]{2019PhRvD..99f3012M}
\bibinfo{author}{{Murase}, K.}, \& \bibinfo{author}{{Fukugita}, M.} (\bibinfo{year}{2019}).
\newblock \bibinfo{title}{{Energetics of high-energy cosmic radiations}}.
\newblock {\it \bibinfo{journal}{\prd}\/},  {\it \bibinfo{volume}{99}\/}\bibinfo{issue}{(6)}, \bibinfo{pages}{063012}. \DOIprefix\doi{10.1103/PhysRevD.99.063012}. \href{http://arxiv.org/abs/1806.04194}{\tt arXiv:1806.04194}.
\bibitem[{{Neronov} et~al.(2025){Neronov}, {Oikonomou} \& {Semikoz}}]{2025PhRvD.111j3025N}
\bibinfo{author}{{Neronov}, A.}, \bibinfo{author}{{Oikonomou}, F.}, \& \bibinfo{author}{{Semikoz}, D.} (\bibinfo{year}{2025}).
\newblock \bibinfo{title}{{Multimessenger signature of cosmic rays from the microquasar V4641 Sgr propagating along a Galactic magnetic field line}}.
\newblock {\it \bibinfo{journal}{\prd}\/},  {\it \bibinfo{volume}{111}\/}\bibinfo{issue}{(10)}, \bibinfo{pages}{103025}. \DOIprefix\doi{10.1103/PhysRevD.111.103025}. \href{http://arxiv.org/abs/2410.17608}{\tt arXiv:2410.17608}.
\bibitem[{{Parra} et~al.(2025){Parra}, {Shidatsu}, {Tomaru}, {Done}, {Mu{\~n}oz-Darias}, {Armas Padilla}, {Ogawa}, {Marino}, {Grollimund}, {Corbel}, {De la Fuente}, {Cheng}, {D{\'\i}az Trigo}, {Fender}, {Isogai}, {Kobayashi}, {Motta}, {Murata}, {Negoro}, {Safi-Harb}, {Suzuki}, {Tsuji}, {Ueda}, {Zhang}, {Zhang} \& {Zhang}}]{V4641_wind_XRISM_2025arXiv250817541P}
\bibinfo{author}{{Parra}, M.}, \bibinfo{author}{{Shidatsu}, M.}, \bibinfo{author}{{Tomaru}, R.} et~al.(\bibinfo{year}{2025}).
\newblock \bibinfo{title}{{XRISM reveals a variable, multi-phase outflow-inflow structure during the X-ray obscured 2024 outburst of the black hole transient V4641 Sgr}}.
\newblock {\it \bibinfo{journal}{arXiv e-prints}\/},  (p. \bibinfo{pages}{arXiv:2508.17541}). \DOIprefix\doi{10.48550/arXiv.2508.17541}. \href{http://arxiv.org/abs/2508.17541}{\tt arXiv:2508.17541}.
\bibitem[{{Pasquevich} et~al.(2026){Pasquevich}, {Romero} \& {Reynoso}}]{2026APh...17703214P}
\bibinfo{author}{{Pasquevich}, L.~M.}, \bibinfo{author}{{Romero}, G.~E.}, \& \bibinfo{author}{{Reynoso}, M.~M.} (\bibinfo{year}{2026}).
\newblock \bibinfo{title}{{Neutrinos from hidden ultraluminous X-ray sources in the Galaxy}}.
\newblock {\it \bibinfo{journal}{Astroparticle Physics}\/},  {\it \bibinfo{volume}{177}\/}, \bibinfo{pages}{103214}. \DOIprefix\doi{10.1016/j.astropartphys.2026.103214}. \href{http://arxiv.org/abs/2601.13378}{\tt arXiv:2601.13378}.
\bibitem[{{Pe'er}(2014)}]{2014SSRv..183..371P}
\bibinfo{author}{{Pe'er}, A.} (\bibinfo{year}{2014}).
\newblock \bibinfo{title}{{Energetic and Broad Band Spectral Distribution of Emission from Astronomical Jets}}.
\newblock {\it \bibinfo{journal}{\ssr}\/},  {\it \bibinfo{volume}{183}\/}\bibinfo{issue}{(1-4)}, \bibinfo{pages}{371--403}. \DOIprefix\doi{10.1007/s11214-013-0001-y}. \href{http://arxiv.org/abs/1306.1355}{\tt arXiv:1306.1355}.
\bibitem[{{Peretti} et~al.(2026){Peretti}, {Amato}, {Cerri}, {Morlino}, {Perfetta Pullano} \& {Recchia}}]{2026arXiv260316647P}
\bibinfo{author}{{Peretti}, E.}, \bibinfo{author}{{Amato}, E.}, \bibinfo{author}{{Cerri}, S.~S.}, \bibinfo{author}{{Morlino}, G.}, \bibinfo{author}{{Perfetta Pullano}, L.}, \& \bibinfo{author}{{Recchia}, S.} (\bibinfo{year}{2026}).
\newblock \bibinfo{title}{{Particle acceleration at recollimation shocks in sub-relativistic jets A model for jets in Seyfert Galaxies, Microquasars and Protostellar Systems}}.
\newblock {\it \bibinfo{journal}{arXiv e-prints}\/},  (p. \bibinfo{pages}{arXiv:2603.16647}). \href{http://arxiv.org/abs/2603.16647}{\tt arXiv:2603.16647}.
\bibitem[{{Peretti} et~al.(2025){Peretti}, {Petropoulou}, {Vasilopoulos} \& {Gabici}}]{2024arXiv241108762P}
\bibinfo{author}{{Peretti}, E.}, \bibinfo{author}{{Petropoulou}, M.}, \bibinfo{author}{{Vasilopoulos}, G.}, \& \bibinfo{author}{{Gabici}, S.} (\bibinfo{year}{2025}).
\newblock \bibinfo{title}{{Particle acceleration and multi-messenger radiation from ultra-luminous X-ray sources: A new class of Galactic PeVatrons}}.
\newblock {\it \bibinfo{journal}{\aap}\/},  {\it \bibinfo{volume}{698}\/}, \bibinfo{pages}{A188}. \DOIprefix\doi{10.1051/0004-6361/202452987}. \href{http://arxiv.org/abs/2411.08762}{\tt arXiv:2411.08762}.
\bibitem[{{Poutanen} et~al.(2007){Poutanen}, {Lipunova}, {Fabrika}, {Butkevich} \& {Abolmasov}}]{2007MNRAS.377.1187P}
\bibinfo{author}{{Poutanen}, J.}, \bibinfo{author}{{Lipunova}, G.}, \bibinfo{author}{{Fabrika}, S.}, \bibinfo{author}{{Butkevich}, A.~G.}, \& \bibinfo{author}{{Abolmasov}, P.} (\bibinfo{year}{2007}).
\newblock \bibinfo{title}{{Supercritically accreting stellar mass black holes as ultraluminous X-ray sources}}.
\newblock {\it \bibinfo{journal}{\mnras}\/},  {\it \bibinfo{volume}{377}\/}\bibinfo{issue}{(3)}, \bibinfo{pages}{1187--1194}. \DOIprefix\doi{10.1111/j.1365-2966.2007.11668.x}. \href{http://arxiv.org/abs/astro-ph/0609274}{\tt arXiv:astro-ph/0609274}.
\bibitem[{{Predehl} et~al.(2021){Predehl}, {Andritschke}, {Arefiev}, {Babyshkin}, {Batanov}, {Becker}, {B{\"o}hringer}, {Bogomolov}, {Boller}, {Borm}, {Bornemann}, {Br{\"a}uninger}, {Br{\"u}ggen}, {Brunner}, {Brusa}, {Bulbul}, {Buntov}, {Burwitz}, {Burkert}, {Clerc}, {Churazov}, {Coutinho}, {Dauser}, {Dennerl}, {Doroshenko}, {Eder}, {Emberger}, {Eraerds}, {Finoguenov}, {Freyberg}, {Friedrich}, {Friedrich}, {F{\"u}rmetz}, {Georgakakis}, {Gilfanov}, {Granato}, {Grossberger}, {Gueguen}, {Gureev}, {Haberl}, {H{\"a}lker}, {Hartner}, {Hasinger}, {Huber}, {Ji}, {Kienlin}, {Kink}, {Korotkov}, {Kreykenbohm}, {Lamer}, {Lomakin}, {Lapshov}, {Liu}, {Maitra}, {Meidinger}, {Menz}, {Merloni}, {Mernik}, {Mican}, {Mohr}, {M{\"u}ller}, {Nandra}, {Nazarov}, {Pacaud}, {Pavlinsky}, {Perinati}, {Pfeffermann}, {Pietschner}, {Ramos-Ceja}, {Rau}, {Reiffers}, {Reiprich}, {Robrade}, {Salvato}, {Sanders}, {Santangelo}, {Sasaki}, {Scheuerle}, {Schmid}, {Schmitt}, {Schwope}, {Shirshakov}, {Steinmetz}, {Stewart}, {Str{\"u}der}, {Sunyaev},
  {Tenzer}, {Tiedemann}, {Tr{\"u}mper}, {Voron}, {Weber}, {Wilms} \& {Yaroshenko}}]{2021A&A...647A...1P}
\bibinfo{author}{{Predehl}, P.}, \bibinfo{author}{{Andritschke}, R.}, \bibinfo{author}{{Arefiev}, V.} et~al. (\bibinfo{year}{2021}).
\newblock \bibinfo{title}{{The eROSITA X-ray telescope on SRG}}.
\newblock {\it \bibinfo{journal}{\aap}\/},  {\it \bibinfo{volume}{647}\/}, \bibinfo{pages}{A1}. \DOIprefix\doi{10.1051/0004-6361/202039313}. \href{http://arxiv.org/abs/2010.03477}{\tt arXiv:2010.03477}.
\bibitem[{{Revnivtsev} et~al.(2002{\natexlab{a}}){Revnivtsev}, {Gilfanov}, {Churazov} \& {Sunyaev}}]{2002A&A...391.1013R}
\bibinfo{author}{{Revnivtsev}, M.}, \bibinfo{author}{{Gilfanov}, M.}, \bibinfo{author}{{Churazov}, E.}, \& \bibinfo{author}{{Sunyaev}, R.} (\bibinfo{year}{2002}{\natexlab{a}}).
\newblock \bibinfo{title}{{Super-Eddington outburst of V4641 Sgr}}.
\newblock {\it \bibinfo{journal}{\aap}\/},  {\it \bibinfo{volume}{391}\/}, \bibinfo{pages}{1013--1022}. \DOIprefix\doi{10.1051/0004-6361:20020865}. \href{http://arxiv.org/abs/astro-ph/0204132}{\tt arXiv:astro-ph/0204132}.
\bibitem[{{Revnivtsev} et~al.(2002{\natexlab{b}}){Revnivtsev}, {Sunyaev}, {Gilfanov} \& {Churazov}}]{2002A&A...385..904R}
\bibinfo{author}{{Revnivtsev}, M.}, \bibinfo{author}{{Sunyaev}, R.}, \bibinfo{author}{{Gilfanov}, M.}, \& \bibinfo{author}{{Churazov}, E.} (\bibinfo{year}{2002}{\natexlab{b}}).
\newblock \bibinfo{title}{{V4641Sgr - A super-Eddington source enshrouded by an extended envelope}}.
\newblock {\it \bibinfo{journal}{\aap}\/},  {\it \bibinfo{volume}{385}\/}, \bibinfo{pages}{904--908}. \DOIprefix\doi{10.1051/0004-6361:20020189}. \href{http://arxiv.org/abs/astro-ph/0109269}{\tt arXiv:astro-ph/0109269}.
\bibitem[{{Romero} et~al.(2017){Romero}, {Boettcher}, {Markoff} \& {Tavecchio}}]{2017SSRv..207....5R}
\bibinfo{author}{{Romero}, G.~E.}, \bibinfo{author}{{Boettcher}, M.}, \bibinfo{author}{{Markoff}, S.}, \& \bibinfo{author}{{Tavecchio}, F.} (\bibinfo{year}{2017}).
\newblock \bibinfo{title}{{Relativistic Jets in Active Galactic Nuclei and Microquasars}}.
\newblock {\it \bibinfo{journal}{\ssr}\/},  {\it \bibinfo{volume}{207}\/}\bibinfo{issue}{(1-4)}, \bibinfo{pages}{5--61}. \DOIprefix\doi{10.1007/s11214-016-0328-2}. \href{http://arxiv.org/abs/1611.09507}{\tt arXiv:1611.09507}.
\bibitem[{{Safi-Harb} et~al.(2022){Safi-Harb}, {Mac Intyre}, {Zhang}, {Pope}, {Zhang}, {Saffold}, {Mori}, {Gotthelf}, {Aharonian}, {Band}, {Braun}, {Fang}, {Hailey}, {Nynka} \& {Rho}}]{Safi-Harb2022}
\bibinfo{author}{{Safi-Harb}, S.}, \bibinfo{author}{{Mac Intyre}, B.}, \bibinfo{author}{{Zhang}, S.} et~al. (\bibinfo{year}{2022}).
\newblock \bibinfo{title}{{Hard X-Ray Emission from the Eastern Jet of SS 433 Powering the W50 ``Manatee'' Nebula: Evidence for Particle Reacceleration}}.
\newblock {\it \bibinfo{journal}{\apj}\/},  {\it \bibinfo{volume}{935}\/}\bibinfo{issue}{(2)}, \bibinfo{pages}{163}. \DOIprefix\doi{10.3847/1538-4357/ac7c05}. \href{http://arxiv.org/abs/2207.00573}{\tt arXiv:2207.00573}.
\bibitem[{{Shaw} et~al.(2022){Shaw}, {Miller}, {Grinberg}, {Buisson}, {Heinke}, {Plotkin}, {Tomsick}, {Bahramian}, {Gandhi} \& {Sivakoff}}]{V4641_wind_chandra2022MNRAS.516..124S}
\bibinfo{author}{{Shaw}, A.~W.}, \bibinfo{author}{{Miller}, J.~M.}, \bibinfo{author}{{Grinberg}, V.} et~al.(\bibinfo{year}{2022}).
\newblock \bibinfo{title}{{High resolution X-ray spectroscopy of V4641 Sgr during its 2020 outburst}}.
\newblock {\it \bibinfo{journal}{\mnras}\/},  {\it \bibinfo{volume}{516}\/}\bibinfo{issue}{(1)}, \bibinfo{pages}{124--137}. \DOIprefix\doi{10.1093/mnras/stac2213}. \href{http://arxiv.org/abs/2208.01732}{\tt arXiv:2208.01732}.
\bibitem[{{Sunyaev} et~al.(2021){Sunyaev}, {Arefiev}, {Babyshkin}, {Bogomolov}, {Borisov}, {Buntov}, {Brunner}, {Burenin}, {Churazov}, {Coutinho}, {Eder}, {Eismont}, {Freyberg}, {Gilfanov}, {Gureyev}, {Hasinger}, {Khabibullin}, {Kolmykov}, {Komovkin}, {Krivonos}, {Lapshov}, {Levin}, {Lomakin}, {Lutovinov}, {Medvedev}, {Merloni}, {Mernik}, {Mikhailov}, {Molodtsov}, {Mzhelsky}, {M{\"u}ller}, {Nandra}, {Nazarov}, {Pavlinsky}, {Poghodin}, {Predehl}, {Robrade}, {Sazonov}, {Scheuerle}, {Shirshakov}, {Tkachenko} \& {Voron}}]{2021A&A...656A.132S}
\bibinfo{author}{{Sunyaev}, R.}, \bibinfo{author}{{Arefiev}, V.}, \bibinfo{author}{{Babyshkin}, V.} et~al. (\bibinfo{year}{2021}).
\newblock \bibinfo{title}{{SRG X-ray orbital observatory. Its telescopes and first scientific results}}.
\newblock {\it \bibinfo{journal}{\aap}\/},  {\it \bibinfo{volume}{656}\/}, \bibinfo{pages}{A132}. \DOIprefix\doi{10.1051/0004-6361/202141179}. \href{http://arxiv.org/abs/2104.13267}{\tt arXiv:2104.13267}.
\bibitem[{{Sunyaev} et~al.(2026){Sunyaev}, {Khabibullin}, {Churazov}, {Gilfanov}, {Medvedev} \& {Sazonov}}]{2026A&A...707A.278S}
\bibinfo{author}{{Sunyaev}, R.}, \bibinfo{author}{{Khabibullin}, I.}, \bibinfo{author}{{Churazov}, E.}, \bibinfo{author}{{Gilfanov}, M.}, \bibinfo{author}{{Medvedev}, P.}, \& \bibinfo{author}{{Sazonov}, S.} (\bibinfo{year}{2026}).
\newblock \bibinfo{title}{{X-ray panorama of the SS 433/W50 complex by SRG/eROSITA}}.
\newblock {\it \bibinfo{journal}{\aap}\/},  {\it \bibinfo{volume}{707}\/}, \bibinfo{pages}{A278}. \DOIprefix\doi{10.1051/0004-6361/202557726}. \href{http://arxiv.org/abs/2510.14938}{\tt arXiv:2510.14938}.
\bibitem[{{Suzuki} et~al.(2025){Suzuki}, {Tsuji}, {Kanemaru}, {Shidatsu}, {Olivera-Nieto}, {Safi-Harb}, {Kimura}, {de la Fuente}, {Casanova}, {Mori}, {Wang}, {Kato}, {Tateishi}, {Uchiyama}, {Tanaka}, {Uchida}, {Inoue}, {Huang}, {Lemoine-Goumard}, {Miura}, {Ogawa}, {Kobayashi}, {Done}, {Parra}, {D{\'\i}az Trigo}, {Mu{\~n}oz-Darias}, {Armas Padilla}, {Tomaru} \& {Ueda}}]{2025ApJ...978L..20S}
\bibinfo{author}{{Suzuki}, H.}, \bibinfo{author}{{Tsuji}, N.}, \bibinfo{author}{{Kanemaru}, Y.} et~al. (\bibinfo{year}{2025}).
\newblock \bibinfo{title}{{Detection of Extended X-Ray Emission around the PeVatron Microquasar V4641 Sgr with XRISM}}.
\newblock {\it \bibinfo{journal}{\apjl}\/},  {\it \bibinfo{volume}{978}\/}\bibinfo{issue}{(2)}, \bibinfo{pages}{L20}. \DOIprefix\doi{10.3847/2041-8213/ad9d11}. \href{http://arxiv.org/abs/2412.08089}{\tt arXiv:2412.08089}.
\bibitem[{{The LHAASO Collaboration} et~al.(2025{\natexlab{a}}){The LHAASO Collaboration}, {Cao}, {Aharonian}, {Bai}, {Bao}, {Bastieri}, {Bi}, {Bi}, {Bian}, {Blunier}, {Bukevich}, {Cai}, {Cai}, {Cao}, {Cao}, {Chang}, {Chang}, {Chen}, {Chen}, {Chen}, {Chen}, {Chen}, {Chen}, {Chen}, {Chen}, {Chen}, {Chen}, {Chen}, {Chen}, {Chen}, {Chen}, {Chen}, {Chen}, {Chen}, {Cheng}, {Cheng}, {Cheng}, {Cui}, {Cui}, {Cui}, {Cui}, {Dai}, {Dai}, {Dai}, {Danzengluobu}, {Diao}, {Dong}, {Dong}, {Duan}, {Fan}, {Fan}, {Fang}, {Fang}, {Fang}, {Feng}, {Feng}, {Feng}, {Feng}, {Feng}, {Feng}, {Feng}, {Gabici}, {Gao}, {Gao}, {Gao}, {Gao}, {Ge}, {Ge}, {Geng}, {Giacinti}, {Gong}, {Gou}, {Gu}, {Guo}, {Guo}, {Guo}, {Guo}, {Guo}, {Guo}, {Han}, {Hannuksela}, {Hasan}, {He}, {He}, {He}, {He}, {He}, {Hern{\'a}ndez-Cadena}, {Hou}, {Hou}, {Hou}, {Hu}, {Hu}, {Huang}, {Huang}, {Huang}, {Huang}, {Huang}, {Huang} et~al.}]{CygX3_LHAASO25}
\bibinfo{author}{{The LHAASO Collaboration}}, \bibinfo{author}{{Cao}, Z.}, \bibinfo{author}{{Aharonian}, F.}, \bibinfo{author}{{Bai}, Y.~X.} et~al. (\bibinfo{year}{2025}{\natexlab{a}}).
\newblock \bibinfo{title}{{Cygnus X-3: A variable petaelectronvolt gamma-ray source}}.
\newblock {\it \bibinfo{journal}{arXiv e-prints}\/},  (p. \bibinfo{pages}{arXiv:2512.16638}). \DOIprefix\doi{10.48550/arXiv.2512.16638}. \href{http://arxiv.org/abs/2512.16638}{\tt arXiv:2512.16638}.
\bibitem[{{The LHAASO Collaboration} et~al.(2025{\natexlab{b}}){The LHAASO Collaboration}, {Cao}, {Aharonian}, {Bai}, {Bao}, {Bastieri}, {Bi}, {Bi}, {Bian}, {Bukevich}, {Cai}, {Cao}, {Cao}, {Chang}, {Chang}, {Chen}, {Chen}, {Chen}, {Chen}, {Chen}, {Chen}, {Chen}, {Chen}, {Chen}, {Chen}, {Chen}, {Chen}, {Chen}, {Chen}, {Chen}, {Chen}, {Cheng}, {Cheng}, {Chung Chu}, {Cui}, {Cui}, {Cui}, {Cui}, {Dai}, {Dai}, {Dai}, {Luobu}, {Diao}, {Dong}, {Duan}, {Fan}, {Fan}, {Fang}, {Fang}, {Fang}, {Feng}, {Feng}, {Feng}, {Feng}, {Feng}, {Feng}, {Feng}, {Gabici}, {Gao}, {Gao}, {Gao}, {Gao}, {Gao}, {Ge}, {Ge}, {Geng}, {Giacinti}, {Gong}, {Gou}, {Gu}, {Guo}, {Guo}, {Guo}, {Guo}, {Guo}, {Han}, {Hannuksela}, {Hasan}, {He}, {He}, {He}, {He}, {He}, {Hern{\'a}ndez-Cadena}, {Hou}, {Hou}, {Hou}, {Hu}, {Hu}, {Huang}, {Huang}, {Huang}, {Huang}, {Huang}, {Huang}, {Huang}, {Huang}, {Huang}, {Ji} et~al.}]{LHAASO_MQs_2025NSRev..12af496L}
\bibinfo{author}{{The LHAASO Collaboration}}, \bibinfo{author}{{Cao}, Z.}, \bibinfo{author}{{Aharonian}, F.}, \bibinfo{author}{{Bai}, Y.-X.} et~al. (\bibinfo{year}{2025}{\natexlab{b}}).
\newblock \bibinfo{title}{{Ultrahigh-Energy Gamma-ray Emission Associated with Black Hole-Jet Systems}}.
\newblock {\it \bibinfo{journal}{National Science Review}\/},  {\it \bibinfo{volume}{12}\/}\bibinfo{issue}{(12)}, \bibinfo{pages}{nwaf496}. \DOIprefix\doi{10.1093/nsr/nwaf496}. \href{http://arxiv.org/abs/2410.08988}{\tt arXiv:2410.08988}.
\bibitem[{{Tomsick} et~al.(2008){Tomsick}, {Chaty}, {Rodriguez}, {Walter} \& {Kaaret}}]{2008ApJ...685.1143T}
\bibinfo{author}{{Tomsick}, J.~A.}, \bibinfo{author}{{Chaty}, S.}, \bibinfo{author}{{Rodriguez}, J.}, \bibinfo{author}{{Walter}, R.}, \& \bibinfo{author}{{Kaaret}, P.} (\bibinfo{year}{2008}).
\newblock \bibinfo{title}{{Chandra Localizations and Spectra of INTEGRAL Sources in the Galactic Plane}}.
\newblock {\it \bibinfo{journal}{\apj}\/},  {\it \bibinfo{volume}{685}\/}\bibinfo{issue}{(2)}, \bibinfo{pages}{1143--1156}. \DOIprefix\doi{10.1086/591040}. \href{http://arxiv.org/abs/0807.2278}{\tt arXiv:0807.2278}.
\bibitem[{{Vecchiotti} et~al.(2026){Vecchiotti}, {Amato}, {Giacinti}, {Morlino} \& {Peron}}]{2026arXiv260629830V}
\bibinfo{author}{{Vecchiotti}, V.}, \bibinfo{author}{{Amato}, E.}, \bibinfo{author}{{Giacinti}, G.}, \bibinfo{author}{{Morlino}, G.}, \& \bibinfo{author}{{Peron}, G.} (\bibinfo{year}{2026}).
\newblock \bibinfo{title}{{Constraints on Hadronic Emission from Microquasars Detected by LHAASO}}.
\newblock {\it \bibinfo{journal}{arXiv e-prints}\/},  (p. \bibinfo{pages}{arXiv:2606.29830}). \DOIprefix\doi{10.48550/arXiv.2606.29830}. \href{http://arxiv.org/abs/2606.29830}{\tt arXiv:2606.29830}.
\bibitem[{{Wan} et~al.(2026){Wan}, {Wang} \& {Liu}}]{shear2026ApJ...997..128W}
\bibinfo{author}{{Wan}, S.-Y.}, \bibinfo{author}{{Wang}, J.-s.}, \& \bibinfo{author}{{Liu}, R.-Y.} (\bibinfo{year}{2026}).
\newblock \bibinfo{title}{{A Leptonic Interpretation of the Ultra-high-energy Gamma-Ray Emission from V4641 Sgr}}.
\newblock {\it \bibinfo{journal}{\apj}\/},  {\it \bibinfo{volume}{997}\/}\bibinfo{issue}{(1)}, \bibinfo{pages}{128}. \DOIprefix\doi{10.3847/1538-4357/ae246c}. \href{http://arxiv.org/abs/2507.02763}{\tt arXiv:2507.02763}.
\bibitem[{{Zhang} et~al.(2025){Zhang}, {Kimura} \& {Murase}}]{2025PhRvD.112l3015Z}
\bibinfo{author}{{Zhang}, B.~T.}, \bibinfo{author}{{Kimura}, S.~S.}, \& \bibinfo{author}{{Murase}, K.} (\bibinfo{year}{2025}).
\newblock \bibinfo{title}{{Microquasar jet-cocoon systems as PeVatrons}}.
\newblock {\it \bibinfo{journal}{\prd}\/},  {\it \bibinfo{volume}{112}\/}\bibinfo{issue}{(12)}, \bibinfo{pages}{123015}. \DOIprefix\doi{10.1103/p6r1-qg5q}. \href{http://arxiv.org/abs/2506.20193}{\tt arXiv:2506.20193}.
\bibitem[{{Zhao} et~al.(2025){Zhao}, {Li} \& {Torres}}]{Fermi_2025ApJ...984....3Z}
\bibinfo{author}{{Zhao}, Z.}, \bibinfo{author}{{Li}, J.}, \& \bibinfo{author}{{Torres}, D.~F.} (\bibinfo{year}{2025}).
\newblock \bibinfo{title}{{Upper Limits on the Gamma-Ray Emission from the Microquasar V4641 Sgr}}.
\newblock {\it \bibinfo{journal}{\apj}\/},  {\it \bibinfo{volume}{984}\/}\bibinfo{issue}{(1)}, \bibinfo{pages}{3}. \DOIprefix\doi{10.3847/1538-4357/adc39c}. \href{http://arxiv.org/abs/2503.16844}{\tt arXiv:2503.16844}.

\end{thebibliography}

\end{document}